\documentclass[namedreferences]{SolarPhysics}
\usepackage[optionalrh]{spr-sola-addons} 
\usepackage{epsfig}                     
\usepackage{graphicx}                    
\usepackage{color}                       
\usepackage{url}                         

\renewcommand{\vec}[1]{{\mathbf #1}}
\begin{document}

\begin{article}

\begin{opening}

\title{Magnetic Energy Storage and Current Density Distributions for Different
 Force-Free Models}

%
\author{S.~\surname{R\'egnier}
       }

%
\runningauthor{S. R{\'e}gnier}
\runningtitle{Magnetic Energy and Current Density}

%
  \institute{$^{1}$ Jeremiah Horrocks Institute, University of Central Lancashire,
  	Preston, Lancashire, PR1 2HE, UK\\
                     email: \url{SRegnier@uclan.ac.uk} 
             }

\begin{abstract}

In the last decades, force-free-field modelling has been used extensively to
describe the coronal magnetic field and to better understand the physics of
solar eruptions at different scales. Especially the evolution of active regions
has been studied by successive equilibria in which each computed magnetic 
configuration is subject to an evolving photospheric distribution of magnetic
field and/or electric current density. This technique of successive equilibria
has been successful in describing the rate of change of the energetics for
observed active regions. Nevertheless the change in the magnetic configuration
due to the increase/decrease of electric current for different force-free models
(potential, linear and nonlinear force-free fields) has never been studied in
detail before. Here we focus especially on the evolution of the free magnetic
energy, the location of the excess of energy, and the distribution of electric
currents in the corona. For this purpose, we use an idealised active region
characterised by four main polarities and a satellite polarity allowing us to
specify a complex topology and sheared arcades to the coronal magnetic field but
no twisted flux bundles. We investigate the changes in the geometry and
connectivity of field lines, the magnetic energy and current density content as
well as the evolution of null points. Increasing the photospheric current
density in the magnetic configuration does not dramatically change the
energy-storage processes within the active region even if the magnetic topology
is slightly modified. We conclude that for reasonable values of the photospheric
current density (the force-free parameter $\alpha < 0.25$ Mm$^{-1}$), the
magnetic configurations studied do change but not dramatically: {\em i}) the original
null point stays nearly at the same location, {\em ii}) the field-line geometry and
connectivity are slightly modified, {\em iii}) even if the free magnetic energy is
significantly increased, the energy storage happens at the same location. This
extensive study of different force-free models for a simple magnetic
configuration shows that some topological elements of an observed active region,
such as null points, can be reproduced with confidence only by considering the
potential-field approximation. This study is a preliminary work aiming at
understanding the effects of electric currents generated by characteristic
photospheric motions on the structure and evolution of the coronal magnetic
field.  

\end{abstract}

%

\end{opening}


\section{Introduction}
\label{sec:intro}

One key unsolved issue in solar physics is the generation and effects of
electric currents from photospheric motions of the stressed and sheared coronal
magnetic field. With the development of reliable techniques such as coronal
magnetic-field extrapolations based on complex distributions of photospheric
currents  (see {\em e.g.}, reviews by \citeauthor{reg07b}, \citeyear{reg07b}; 
\citeauthor{wie08}, \citeyear{wie08}), it is
important to understand how a modelled magnetic-field configuration is subject
to change due to slight modifications of the photospheric-current distribution
and thus the difference between several force-free assumptions using the same
boundary conditions. The main aim is to understand these changes in the geometry
and topology of field lines assuming that the magnetic field is in a force-free
equilibrium. In \inlinecite{reg07}, we have performed a first comparison between
four different active regions with different behaviours due to the complex
distribution of the photospheric field and of the electric currents representing
the history of the evolution of the active region. This comparison was done for
the same force-free model, namely the nonlinear force-free field. We showed
that, statistically speaking, the magnetic-field lines are longer and higher in
a nonlinear force-free field compared to the corresponding potential field. To
develop our understanding of the effects of electric currents on magnetic
configurations, \inlinecite{reg09a} compared the behaviour of a simple bipolar
field subject to different distributions of electric currents using different
force-free models. The bipolar field has been studied in terms of
magnetic-energy storage and magnetic-helicity changes. We showed that
the amount of electric currents that can be injected in a magnetic-equilibrium
configuration depends strongly on the spatial distribution of the currents, the
existence of return currents having a stabilising effect on the magnetic
configuration. The next step developed in this article is to understand the
effects of electric currents on a magnetic configuration having predominant
topological elements ({\em i.e.} a null point in the domain of interest).

In the past decades, magnetic topology has become a key ingredient in
understanding the origin of flares in active regions. 

The general definition of magnetic topology concerns the properties of
magnetic-field lines and magnetic-flux surfaces that are invariant under
continuous deformation in plasma conditions satisfying the frozen-in assumption
\cite{low06, low07, jan10a, ber06}. This definition implies that the number of
null points does not change and the connectivity of field lines rigidly anchored
to the boundaries of the domain is also invariant under the above conditions.
However, we focus here on the evolution of topological elements and connectivity
of field lines obtained from force-free models: the field lines are not rigidly
anchored (the models allow for reconnection of field lines) in order to maintain
a force-free state. Since the magnetic topology is not required to be preserved,
we restrict the definition of the magnetic topology for a given equilibrium
state to the ensemble of topological elements forming the magnetic skeleton
\cite{bun96}. Despite an extensive literature on magnetic topology, the
theoretical background in three dimensions was developed only recently (see
review by \opencite{pri00}). The 3D topological elements constituting the
skeleton of a magnetic configuration can be divided into two parts: the true
topology containing null points, separatrix surfaces, and separators, and the
quasi-topology including quasi-separatrix layers and hyperbolic flux tubes (and
the true topology). We focus especially on the location and properties of
null points as a proxy for describing the topology of magnetic configuration.
This study aims at understanding the possible changes of properties and location
of null points subject to the continuous variation of a free parameter in
various force-free models.

It has been proven that the topological elements of a coronal magnetic
configuration are of prime importance to study the onset of flares and coronal
mass ejections (CMEs). For instance, in the classical model of flare, magnetic
reconnection occurs at a null point or in a current sheet formed along a
topological element. Combining observations and coronal-field models,
\citeauthor{aul00} have shown that a powerful flare associated with a CME
involves a coronal null point and a spine field line. This topological study in
conjunction with EUV observations supports the breakout model \cite{ant99} as a
triggering mechanism for this particular event. Recently, \inlinecite{zhao08}
have derived the skeleton of an active region and its temporal evolution before and
after an eruptive event. The authors have found several coronal null points in a
quasi-force-free field configuration. Unfortunately, the null points found by
\inlinecite{zhao08} did not satisfy the properties of null points for force-free
fields and for divergence-free fields (see Appendix A). Other topological
studies have been carried out to better understand the release of magnetic
energy and the reconnection processes in active region evolution
\cite{dem94,den05,bar05,reg06,li06,luo07,bar07}, during blinkers  \cite{sub08},
and in the quiet Sun \cite{schr02,clo04,reg08a} 

Our work has been motivated by two earlier articles on this topic by
\inlinecite{dem94} and \inlinecite{hud99}. \inlinecite{dem94} have compared the topology of
potential and linear force-free fields for a quadrupolar configuration with a
coronal null. They found that the topology is similar for the point-charge model
but they also noticed that for a bipolar model (based on extended sources)
another null can be created in the linear force-free configuration for large
value of the force-free parameter $\alpha$. Based on the point-charge model,
\inlinecite{bun96} reached the same conclusion for different charge distributions.
\inlinecite{hud99} have considered a quadrupolar point charge, symmetric distribution
to study the connectivity of field lines for potential, linear, and nonlinear
force-free fields. They concluded that the topology can be insanely different
from one model to the other. For non-symmetric cases, \inlinecite{bro00} have found
that the topology of a force-free configuration can be similar using the
point-charge model. As a step forward to the understanding of the magnetic
topology of reconstructed coronal fields, we carry out a comparison between
different models of magnetic fields (potential, linear, and nonlinear force-free
fields) for a configuration having a coronal null point assuming a continuous
distribution of magnetic field at the bottom boundary and no symmetry. We
describe the changes in the magnetic configurations based on the evolution of
the geometry and connectivity of magnetic field lines and in terms of
distribution of magnetic energy and electric currents. In addition, we study the
properties of null points as a proxy of the complexity of the field, keeping in
mind that the separators and separatrices play an important role in the release
of magnetic energy ({\em e.g.}, \citeauthor{pri05}, \citeyear{pri05}). It is
worth noticing that we only focus on the modelling of magnetic-field equilibria
and their changes through the increase of electric currents, whilst recently
\inlinecite{san11} have described the changes in magnetic fields subject to
characteristic photospheric motions using a magnetohydrodynamic (MHD) approach.
They have found that the topology of a quadrupolar magnetic field remains
stable whatever the perturbations imposed even in the case of strong currents.
The  authors claimed that their results can easily be generalised to more
complex magnetic-field distributions and non-generic and symmetric cases.
Despite their sophisticated MHD approach, they do not study the changes
in magnetic energy or in the connectivity of field lines.   

In Section~\ref{sec:mk_np}, we construct a magnetic configuration with a coronal
null point from which we will reconstruct the different models (see
Section~\ref{sec:model}). We thus analyse the geometry of field lines in
Section~\ref{sec:geo} and their connectivity in Section~\ref{sec:con}. The
change in topology in the different models is discussed in
Section~\ref{sec:topo}. In addition, we study how the magnetic energy
is stored (Section~\ref{sec:nrj}) and the electric currents are concentrated
(Section~\ref{sec:cur}). In Section~\ref{sec:concl}, we discuss the implications
for future topological studies from reconstructed magnetic fields.

\begin{figure}
\centering
\includegraphics[width=.531\linewidth]{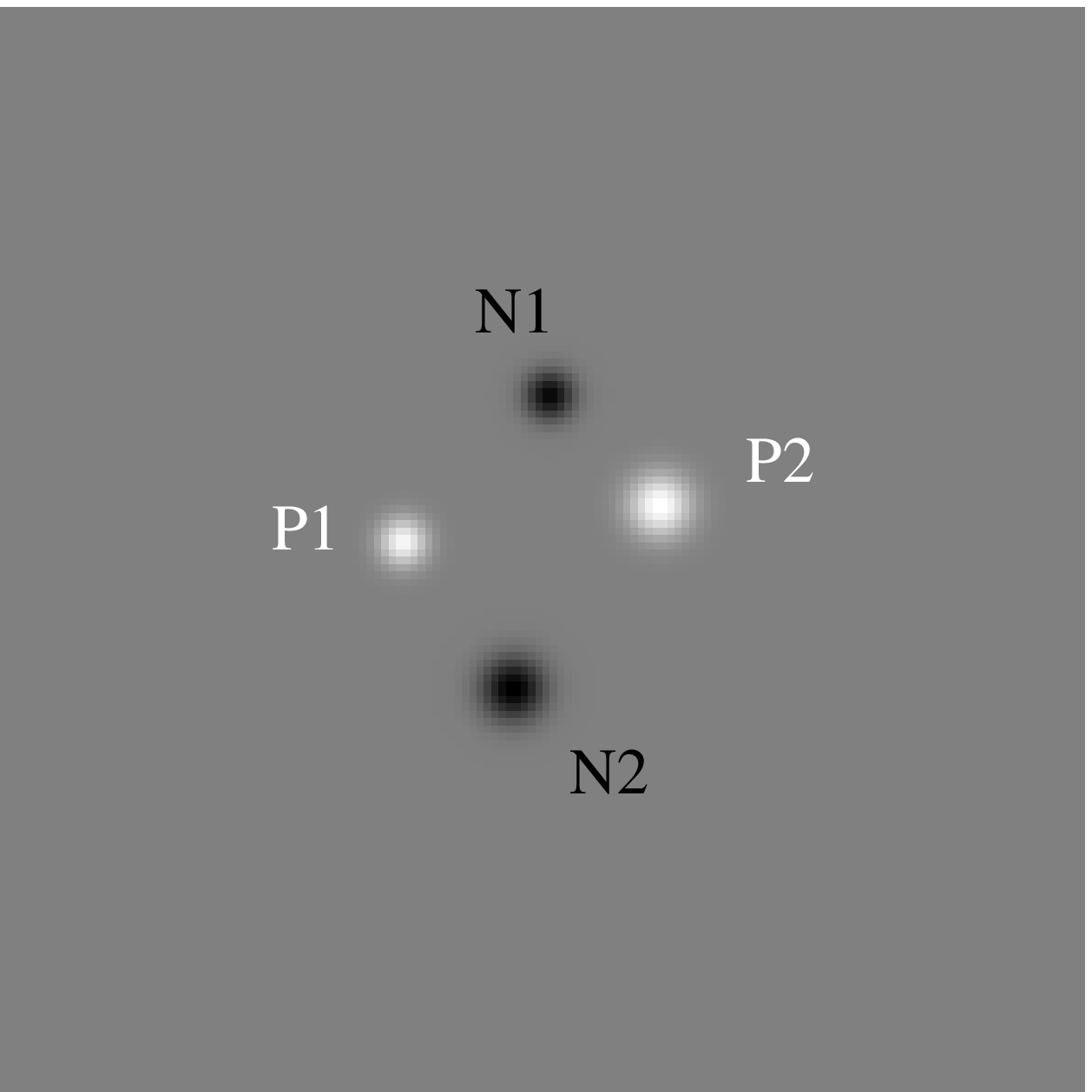}
\includegraphics[width=.459\linewidth, bb= 130 100 790 860,
clip=]{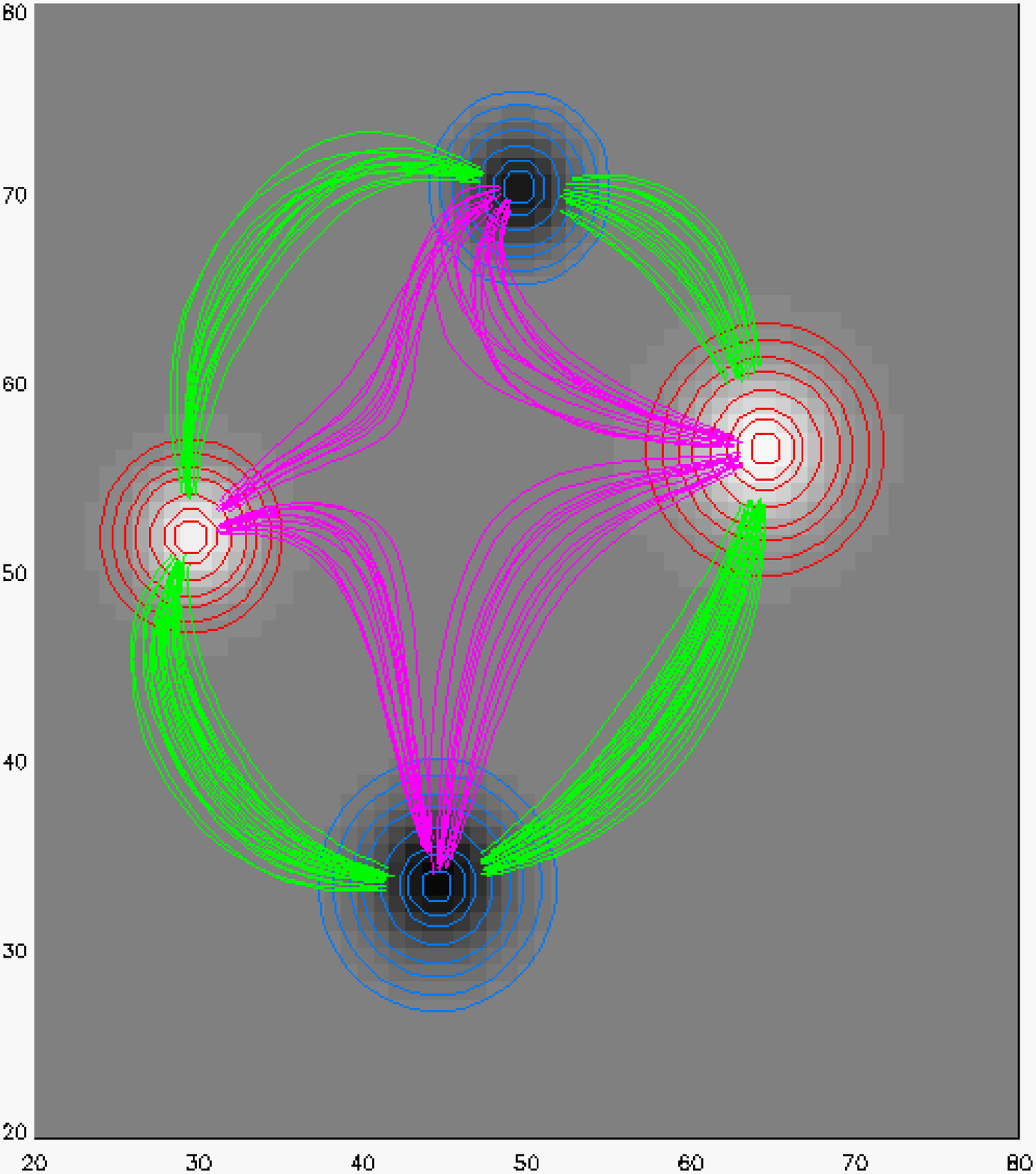}
\caption{(Left) Quadrupolar distribution of the vertical component of the
magnetic field ($B_z$) used as initial field (black and white are negative (N1
and N2) and positive (P1 and P2) polarities). The total magnetic flux is
balanced. (Right) Close-up of a few field lines (red and green) depicting the
geometry and topology of the potential-field configuration. White polarities and
red contours (black polarities and blue contours) are positive ( negative)
values of $B_z$. }
\label{fig:bz_quadru}
\end{figure}


\section{Constructing a Coronal Null Point} 

	\subsection{The Potential-Field Distribution} \label{sec:mk_np}

We first build a quadrupolar distribution of the vertical component of the
magnetic field ($B_z$) at the bottom boundary (see Figure~\ref{fig:bz_quadru} left).
The polarities are defined as Gaussian distributions with the field strength at
the centre and the width of the distribution as free parameters: N1 and N2
(P1 and P2) are negative (positive) polarities. To realistically
model a solar active region, we assume that the spatial resolution is of 1 Mm
giving a characteristic size of 140 Mm. The four polarities are placed such that
there is no symmetry. Each polarity has a maximum field strength of 2000 G in
absolute value and a different width of the Gaussian distribution. The total
magnetic flux is balanced. 

From the magnetogram depicted in Figure~\ref{fig:bz_quadru} left, we compute the
potential field in the coronal volume ($\Omega$: 140 pixels$\times$140
pixels$\times$120 pixels) imposing closed boundary conditions on the sides and
top of the computational box. Few magnetic field lines have been selected in
Figure~\ref{fig:bz_quadru} right to show the geometry and topology of the
potential field associated with this quadrupolar distribution. There is no null
point in the potential-field configuration. However, the magnetic configuration
has a topology characterised by quasi-separatrix layers (QSLs) as shown, in
Figure~\ref{fig:bz_quadru} right, by the red field lines dividing the domain in
four distinct regions. The study of the QSLs is beyond the scope of this
article. We note that the simplest magnetic configuration with a topology
is a configuration with three polarities as studied in detail by \inlinecite{bro99}.
We have chosen a quadrupolar configuration in order to confine the null
point inside the strong-field region. 

\begin{figure}
\centering
\includegraphics[width=.49\linewidth]{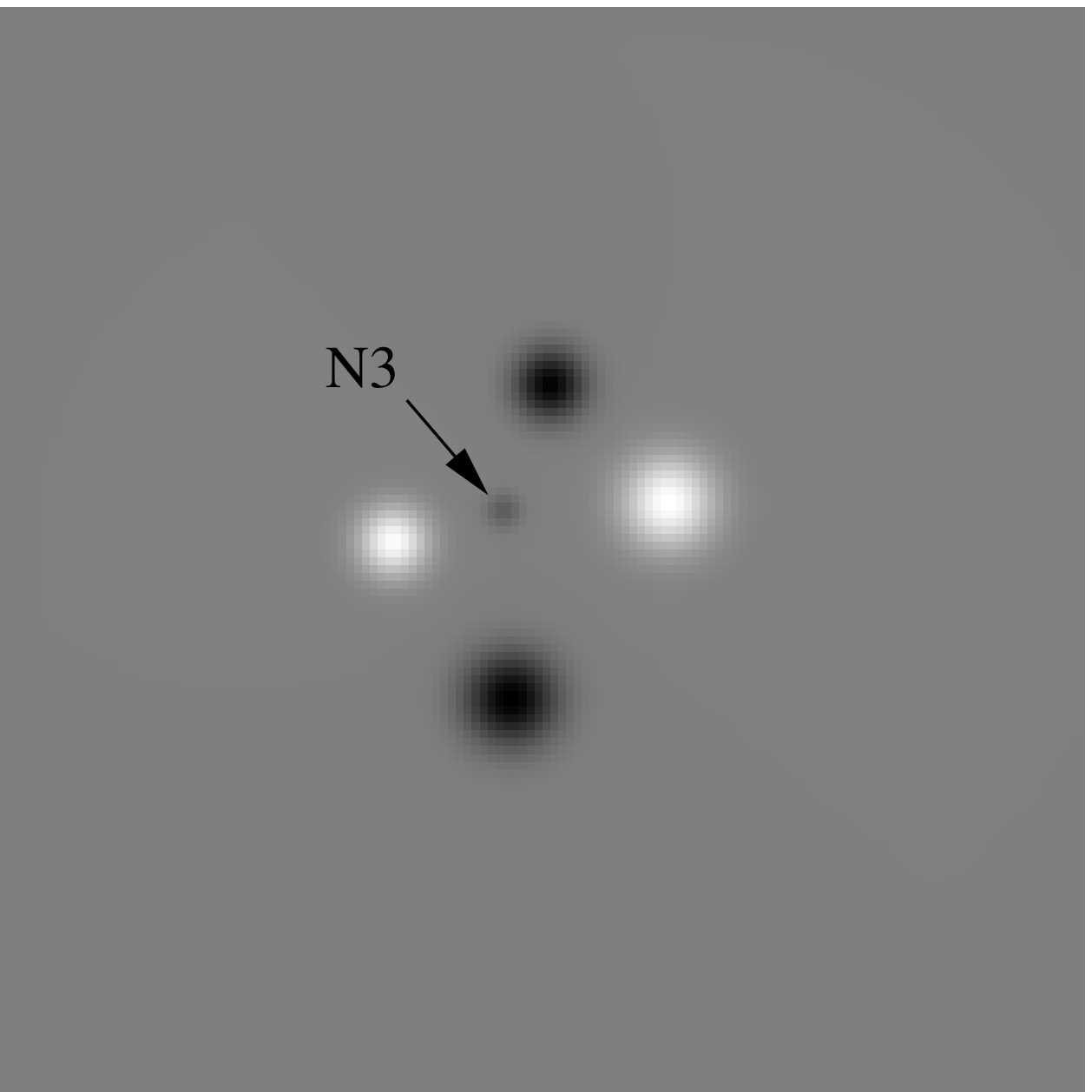}
\includegraphics[width=.49\linewidth]{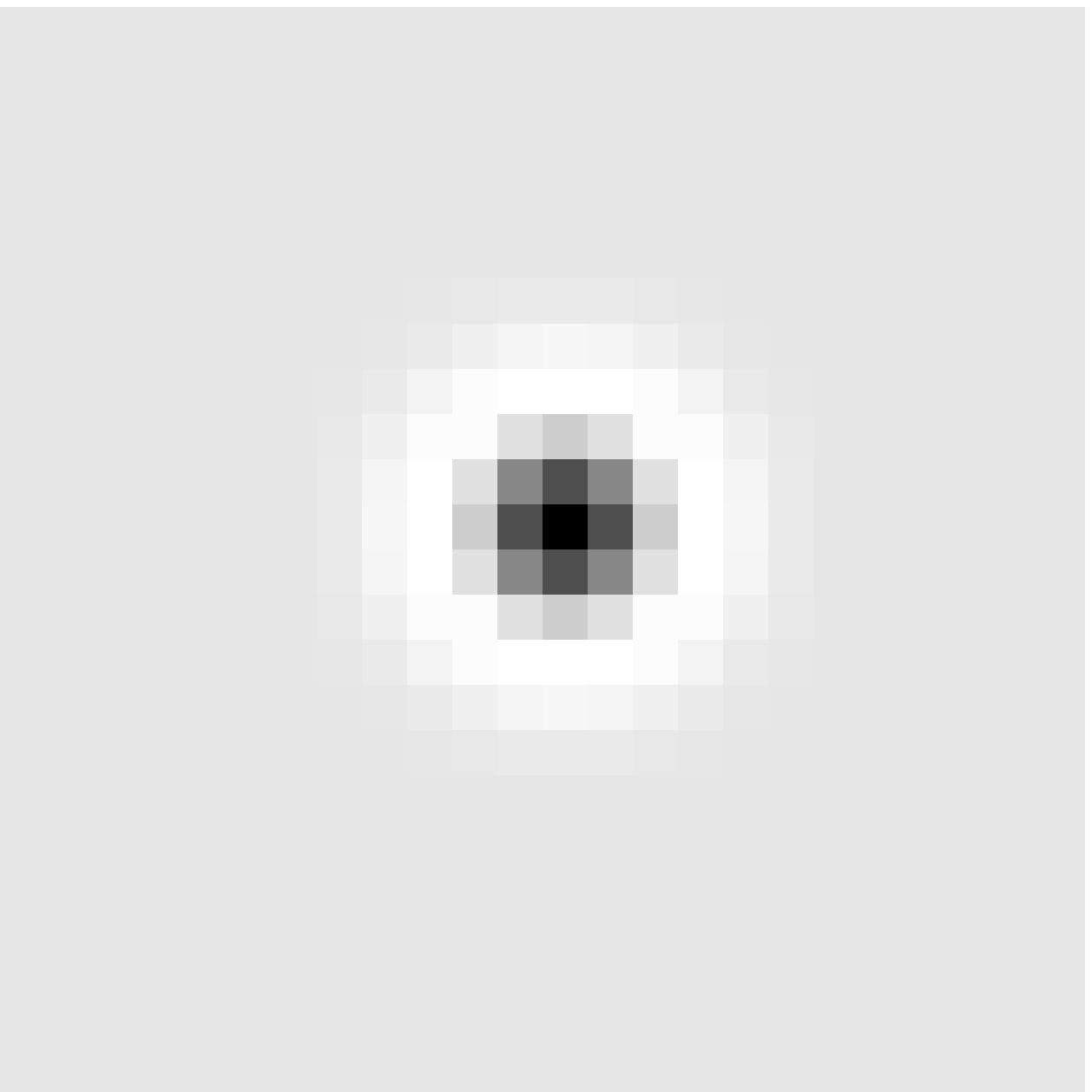}
\caption{(Left) Distribution of the vertical component of the magnetic field at
the bottom boundary (black and white for negative and positive polarities);
(Right) Typical ring distribution of $J_z$ that we impose on the positive
polarities}
\label{fig:jz}
\end{figure}

To create a coronal null point, we emerge a polarity N3 (negative in this
experiment) at the location where there is a field-strength minimum of the
quadrupolar distribution (see Figure~\ref{fig:jz} left). The total magnetic flux
is kept balanced.  We then compute the potential field associated with this new
magnetogram, and with the same boundary conditions as in the quadrupolar case.
The height of the null point depends on the field strength of the polarity N3:
for a maximum field strength of --800 G, the null point is located 6.9 Mm
above the bottom boundary (see also Table~\ref{tab:prop}). In the following
experiment, we will use the vertical component of the magnetic field depicted in
Figure~\ref{fig:jz} left as a boundary condition for the different force-free
models.

	\subsection{The Magnetic-Field Models} \label{sec:model}

\subsubsection*{The Grad--Rubin Algorithm}

The computed linear and nonlinear force-free fields are based on the
\citeauthor{gra58} (\citeyear{gra58}) numerical scheme described by
\citeauthor{ama97} (\citeyear{ama97}, \citeyear{ama99b}) and \inlinecite{ama06}. The same boundary conditions
are used for all the models: the vertical magnetic-field component everywhere
and the distribution of $\alpha$ in one chosen polarity on the bottom boundary,
and closed boundary conditions on the sides and top of the computational box.
Using the same boundary conditions for all models allows us to perform a direct
comparison of the magnetic energy and the topology of the different magnetic
configurations.

The force-free field in the volume above the bottom boundary is thus governed by the following equations:
\begin{equation}
\vec \nabla \times \vec B = \alpha \vec B,
\end{equation}
\begin{equation}
\vec B \cdot \vec \nabla \alpha = 0,
\label{eq:alphaline}
\end{equation}
\begin{equation}
\vec \nabla \cdot \vec B = 0,
\label{eq:divb}
\end{equation}
where $\vec B$ is the magnetic-field vector in the domain $\Omega$ above the
photosphere ($\delta \Omega$), and $\alpha$ is a function of space defined as
the ratio of the vertical current density ($J_{z}$), and the vertical magnetic
field  component ($B_{z}$). From
Equation~(\ref{eq:alphaline}), $\alpha$ is constant along a field line. In terms
of the magnetic field $\vec B$, the Grad--Rubin iterative scheme can be written
as follows:

\begin{equation}
\vec B^{(n)} \cdot \vec \nabla \alpha^{(n)} = 0 \quad \textrm{in} \quad \Omega, 
\end{equation}
\begin{equation}
\alpha^{(n)}|_{\delta \Omega^{\pm}} = h,
\label{eq:bcalpha}
\end{equation}
where $\delta \Omega^{\pm}$ is defined as the domain on the photosphere for which
$B_{z}$ is positive ($+$) or negative ($-$)
and,
\begin{equation}
\vec \nabla \times \vec B^{(n+1)} = \alpha^{(n)} \vec B^{(n)} \quad \textrm{in} \quad
\Omega,
\end{equation}
\begin{equation}
\vec \nabla \cdot \vec B^{(n+1)} = 0 \quad \textrm{in} \quad \Omega, 
\end{equation}
\begin{equation}
B_{z}^{(n+1)}|_{\delta \Omega} = g,
\label{eq:bcbz}
\end{equation}
\begin{equation}
\lim_{|r| \rightarrow \infty}~|\vec B| = 0.
\label{eq:binf}
\end{equation}
The boundary conditions on the photosphere are given by the distribution ($g$) of
$B_{z}$ on $\delta \Omega$ (see Equation~(\ref{eq:bcbz})) and by the
distribution ($h$) of $\alpha$ on $\delta \Omega$ for a given polarity (see
Equation~(\ref{eq:bcalpha})). We also impose that 
\begin{equation}
B_{n} = 0 \quad \textrm{on} \quad \Sigma - \delta \Omega
\label{eq:bcsurf}
\end{equation}
where $\Sigma$ is the surface of the computational box, $n$ refers to the
component normal to the surface. These conditions mean that no field line can
enter or leave the computational box. To ensure the latter condition, we have
chosen a bottom boundary large enough for the magnetic-field strength to tend to
zero near the edges of the field-of-view.

\subsubsection*{Linear Force-Free Fields}

The linear force-free models are based on the Grad--Rubin algorithm where the
distribution of $\alpha$ is a constant. We choose values of $\alpha$ ranging
from --1 to 1 Mm$^{-1}$ with a step $\delta \alpha = 0.02$ Mm$^{-1}$. These
$\alpha$ values correspond to active-region values reported, for instance, by
\inlinecite{lek99a} and computed by assuming that the measured photospheric field is
force-free.

\subsubsection*{Nonlinear Force-Free Fields}

In addition to the vertical component of the magnetic field, the
Grad--Rubin scheme requires a distribution for the current density (or
$\alpha$) in order to derive the nonlinear force-free field. Our choice goes to
the so-called {\em ring} distribution defined by a second-order Hermite
polynomial function as follows:
\begin{equation}
J_z = 2~J_{z0}~[r^2 - C_0]~\exp{\left(-\frac{r^2}{\sigma^2}\right)},
\end{equation} 
where $r$ is measured from the centre of the source and $C_0$ is a constant
that ensures a zero net current. An example of ring distribution is depicted in
Figure~\ref{fig:jz} right. For the sake of comparison, we choose $J_{z0}$
[-20, -10, 10, 20] mA m$^{-2}$, which are characteristic values of the
current density in active regions as has been measured in the photosphere
from vector magnetograms ({\em e.g.}, \citeauthor{lek99a}, \citeyear{lek99a}).
The distribution of $\alpha$ is then given by:
\begin{equation}
\alpha = \frac{\mu_0 J_z}{B_z}.
\end{equation}
In accordance with the Grad--Rubin mathematically well-posed boundary-value
problem, we impose the distribution of $\alpha$ in one chosen polarity (the
positive polarities in this experiment) as boundary condition for the nonlinear
force-free field. We use the same grid and the same side and top boundary
conditions as for the potential and linear force-free fields. 

This particular choice of the vertical current distribution is justified by the
study reported by R\'egnier (2009), which analysed the behaviour of a
simple bipolar field under the assumption of a nonlinear force-free field by
using several distribution of $J_z$ (or $\alpha$). This study showed that the
ring distribution of current gives magnetic configurations that are more stable
and in which a large amount of current can be injected. The conclusion is easily
explained by the stabilising effects of the return currents. Ring-current
distributions have been used before for simulating twisted flux tubes in MHD
models \cite{mag03}.   

\section{Statistical Study of Field-Line Geometry}
\label{sec:geo}


\begin{figure}
\centering
\includegraphics[width=0.49\textwidth]{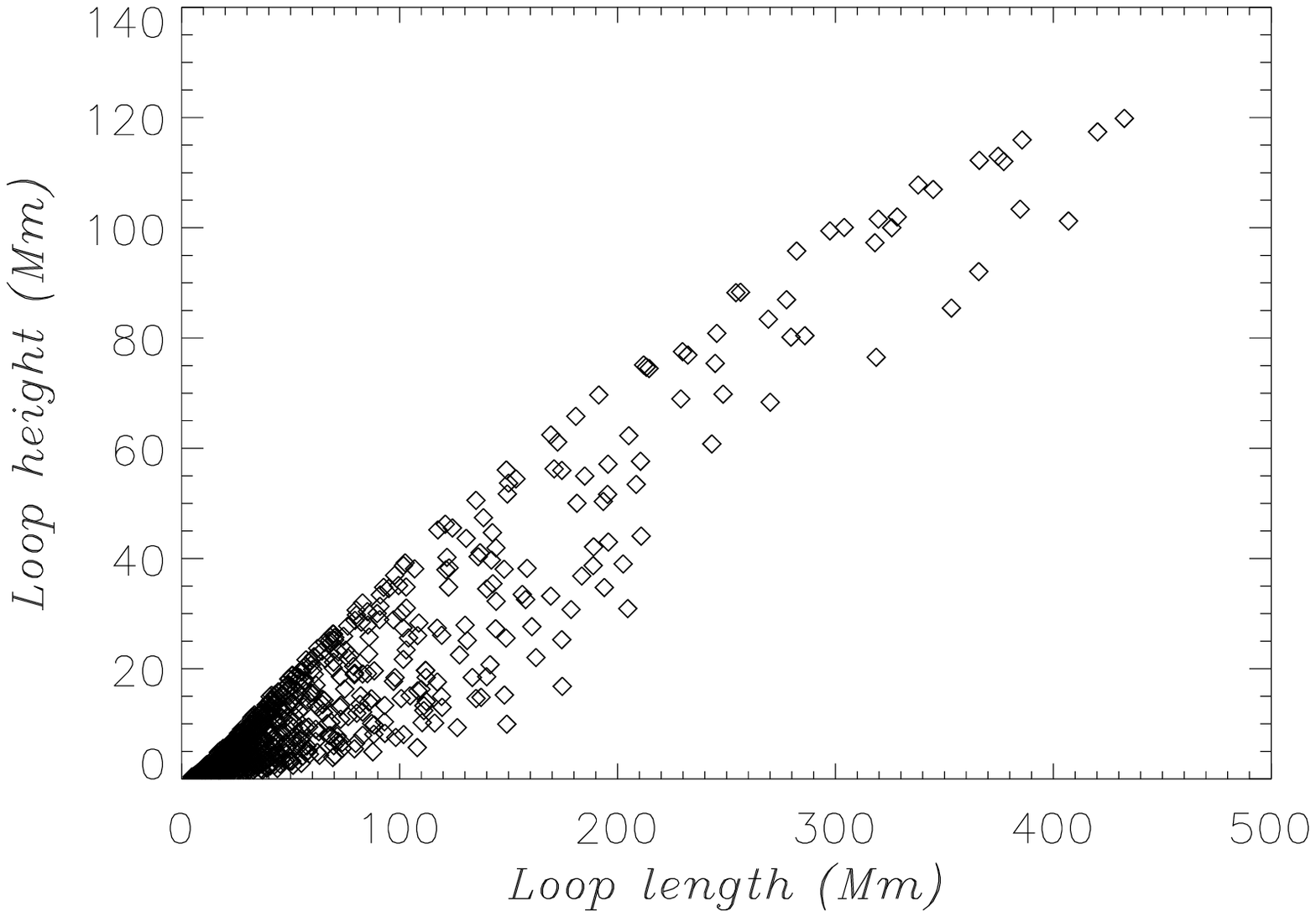}
\includegraphics[width=0.49\textwidth]{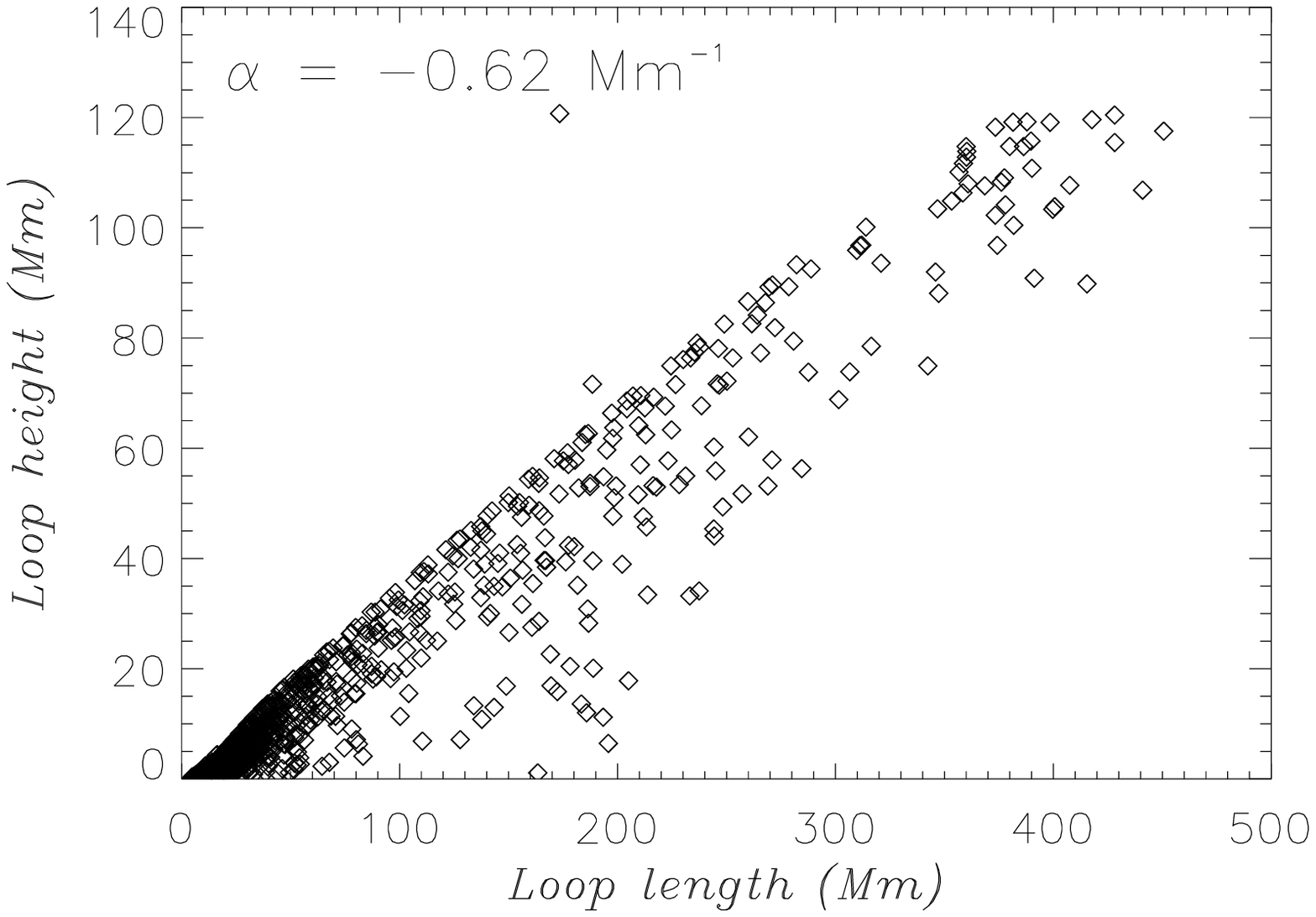}
\includegraphics[width=0.49\textwidth]{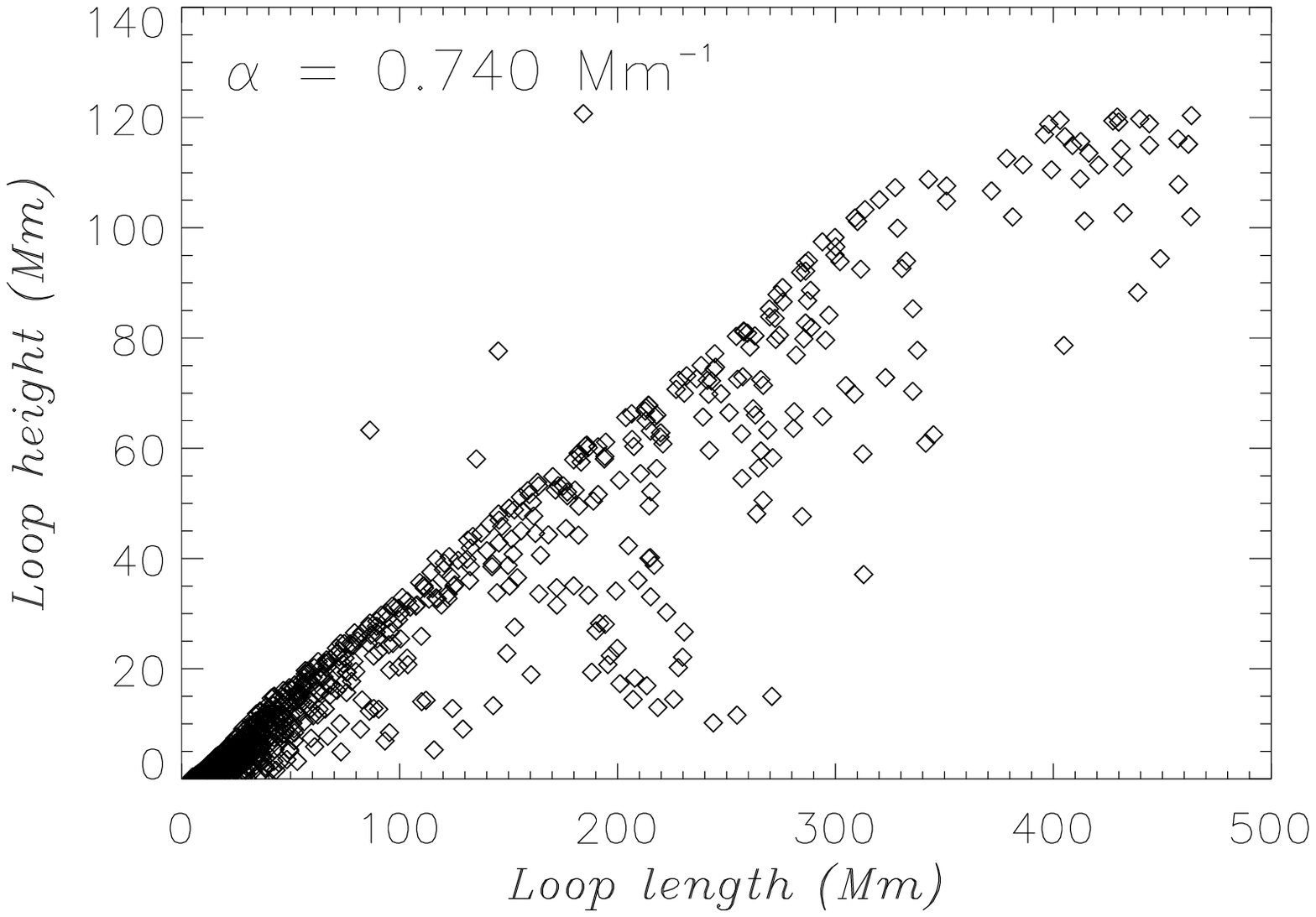}
\includegraphics[width=0.49\textwidth]{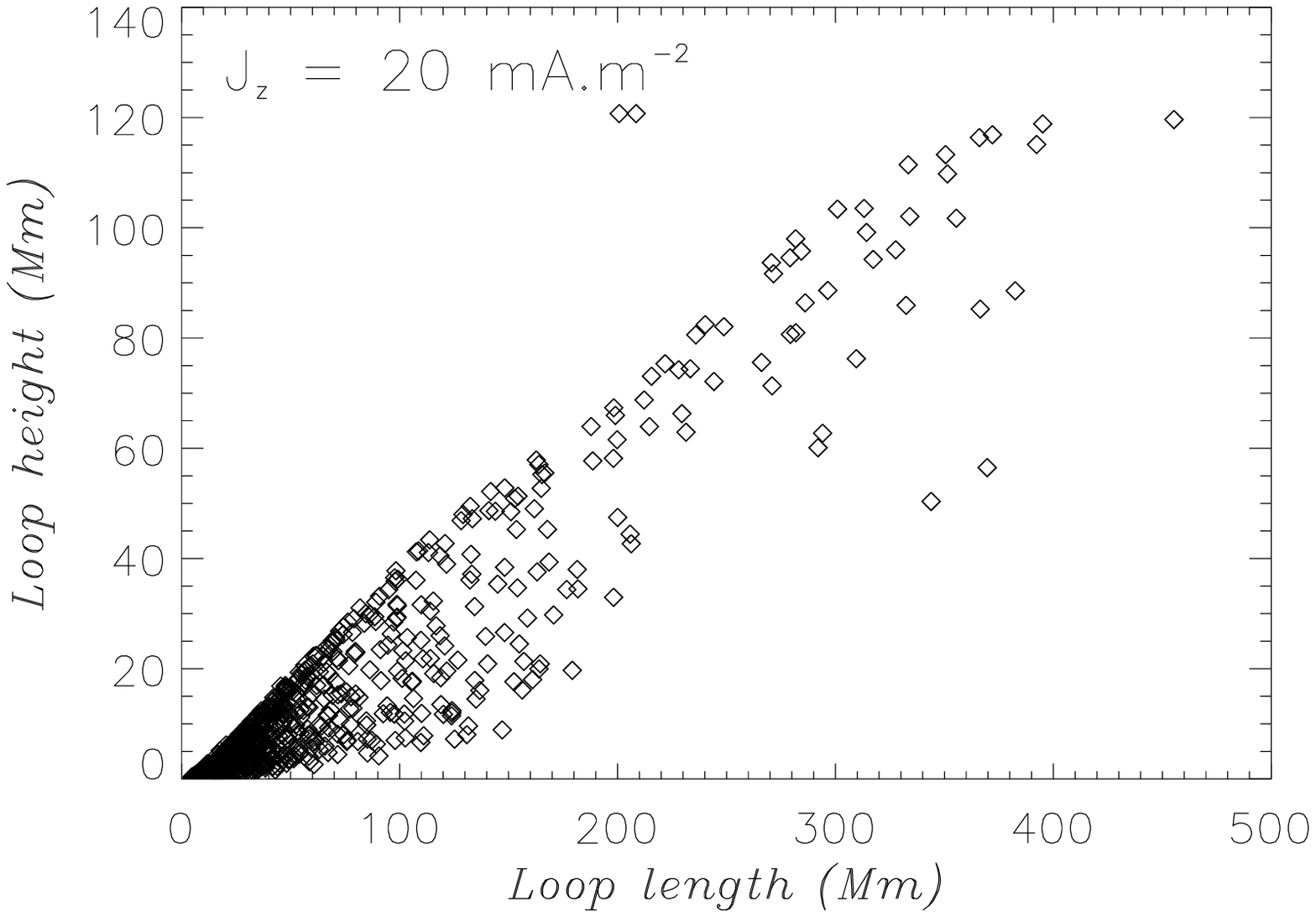}
\caption{Scatter plots of the length and height of selected field lines for (a)
the potential field, (b) and (c) the linear force-free field with $\alpha =
-0.62, 0.74$ Mm$^{-1}$ respectively, and (d) the nonlinear force-free field with
$J_{z0} = 20$ mA m$^{-2}$\,.}
\label{fig:geo}
\end{figure}

To study the differences between the different magnetic configurations obtained
for each model, the first step is to analyse the changes in the geometry of
field lines. So we first select field lines by considering their footpoints with
a field strength above 100 G in absolute value. In the following, we will focus on four
different distributions which are characteristic of this study: the potential
field, two linear force-free field configurations for $\alpha = -0.62, 0.74$
Mm$^{-1}$ and one nonlinear force-free field with $J_{z0} = 20$
mA m$^{-2}$. 

\begin{figure}
\centering
\includegraphics[width=0.49\textwidth]{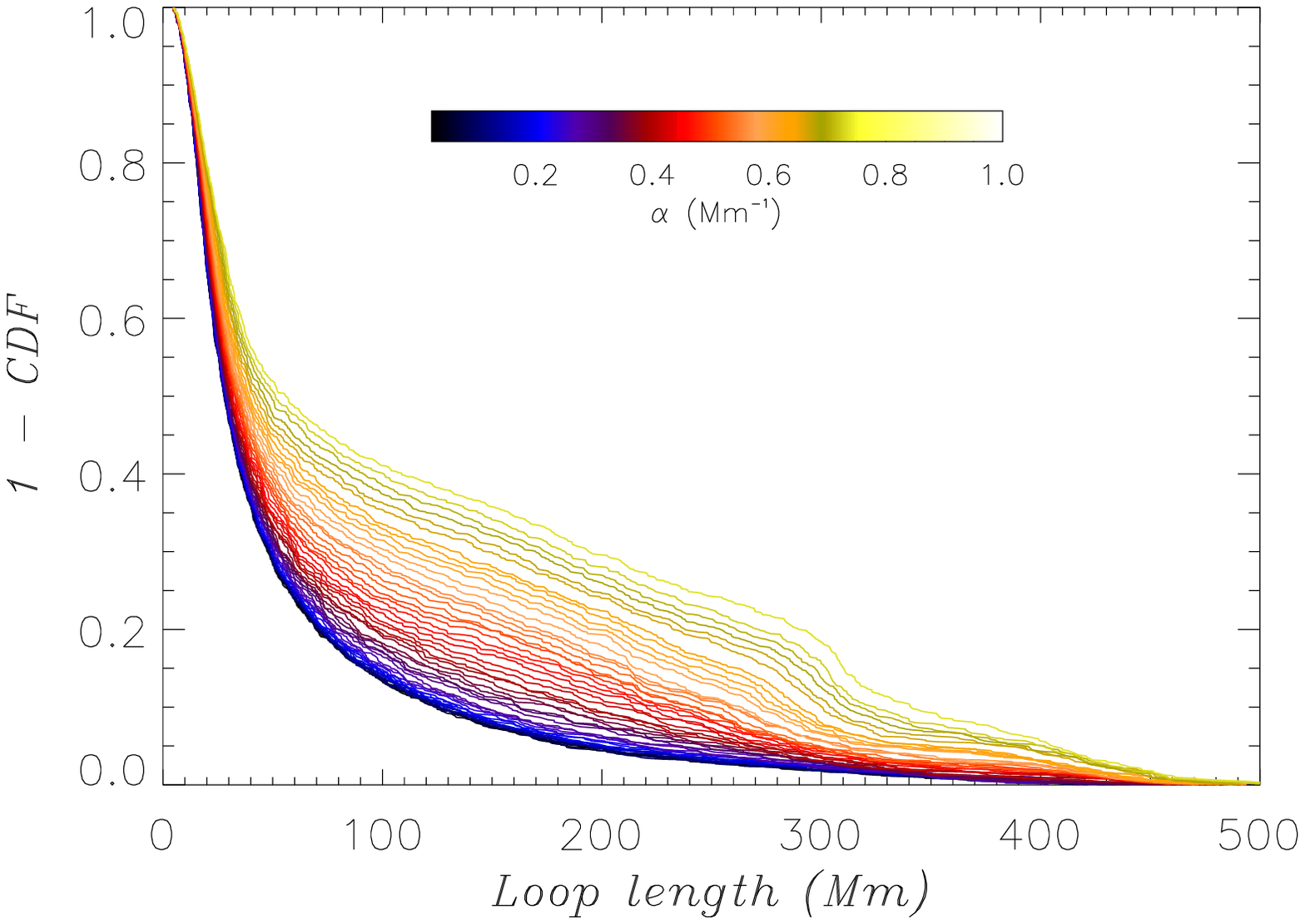}
\includegraphics[width=0.49\textwidth]{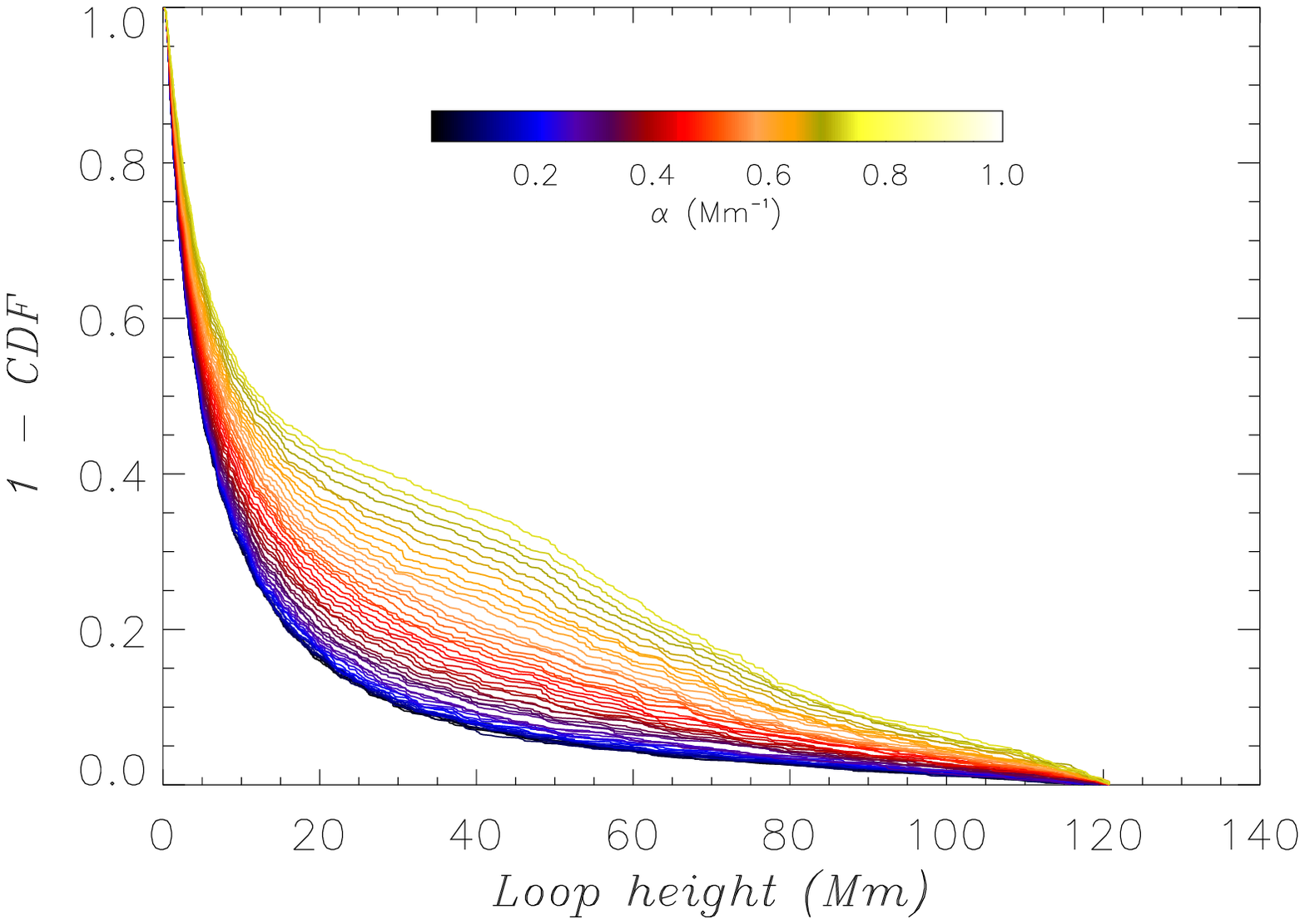}
\caption{Cumulative distribution function (CDF) for the linear force-free models
(coloured curves for the different values of $\alpha$) for the field-line length
(left) and the field line height (right). We only plot the distribution for
positive values of $\alpha$ to avoid confusion.}
\label{fig:geo_lfff}
\end{figure}

In Figure~\ref{fig:geo}, the scatter plots of the length and height of the
selected field lines are shown for the four configurations. It is clear that the
injection of electric currents in a magnetic configuration changes, in a
statistical sense, the geometry
of field lines compared to the potential field but not the same
way for each model. The changes in geometry are more pronounced for the linear
force-free models whilst the geometry remains similar for the nonlinear
force-free field. As mentioned above, the similarity of the nonlinear force-free
model and the potential field and the differences with the linear force-free 
fields are due to the existence of return currents and their stabilising
effects. We now study more closely how the changes in the geometry of field
lines evolve when the current density is increased. In
Figure~\ref{fig:geo_lfff}, we plot the cumulative-distribution functions of the
length (left) and height (right) of selected field lines. We determine that 50\%
of the field lines are shorter than 25 Mm and lower than 5 Mm for the potential
field whilst 50\% of the field lines reaches 50 Mm in length and 15 Mm in height
for $\alpha = 1$ Mm$^{-1}$. The differences between the different linear
force-free models occur for field lines longer than 50 Mm and higher than 10 Mm.
We also notice that the main differences occur for values of $\alpha$ greater
than 0.25 Mm$^{-1}$. The CDFs are consistent with the analysis of
\inlinecite{reg07}: by studying the force-free-field
configurations of four different active regions, the authors have concluded
that, statistically, the field lines in a nonlinear force-free field are longer
and higher than for the corresponding potential field configuration. We will
later refer to values of $\alpha$ less than 0.25 Mm$^{-1}$ as reasonable values
of $\alpha$.

\section{Field-Line Connectivity}
\label{sec:con}

\begin{figure}
\centering
\includegraphics[width=0.49\textwidth]{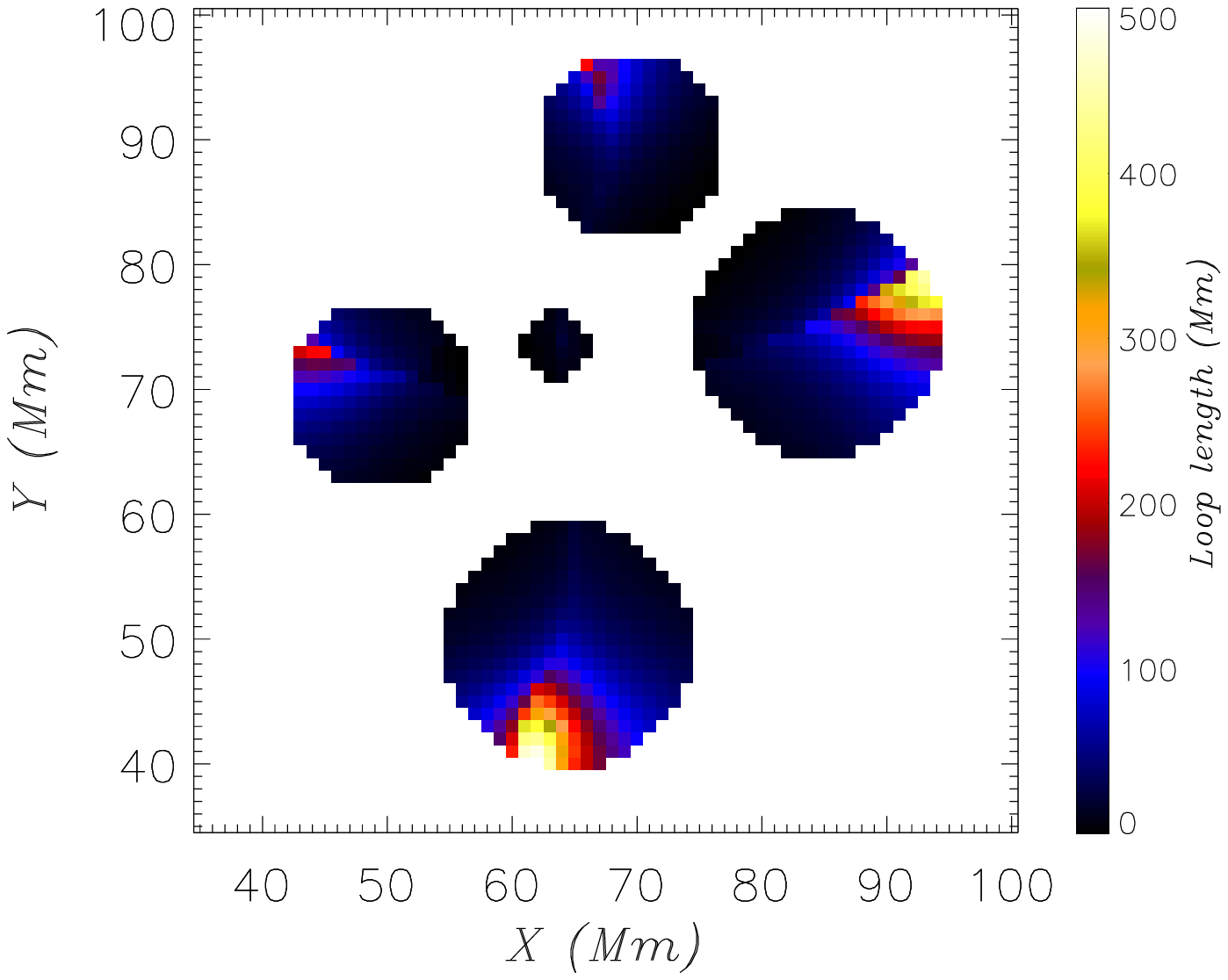}
\includegraphics[width=0.49\textwidth]{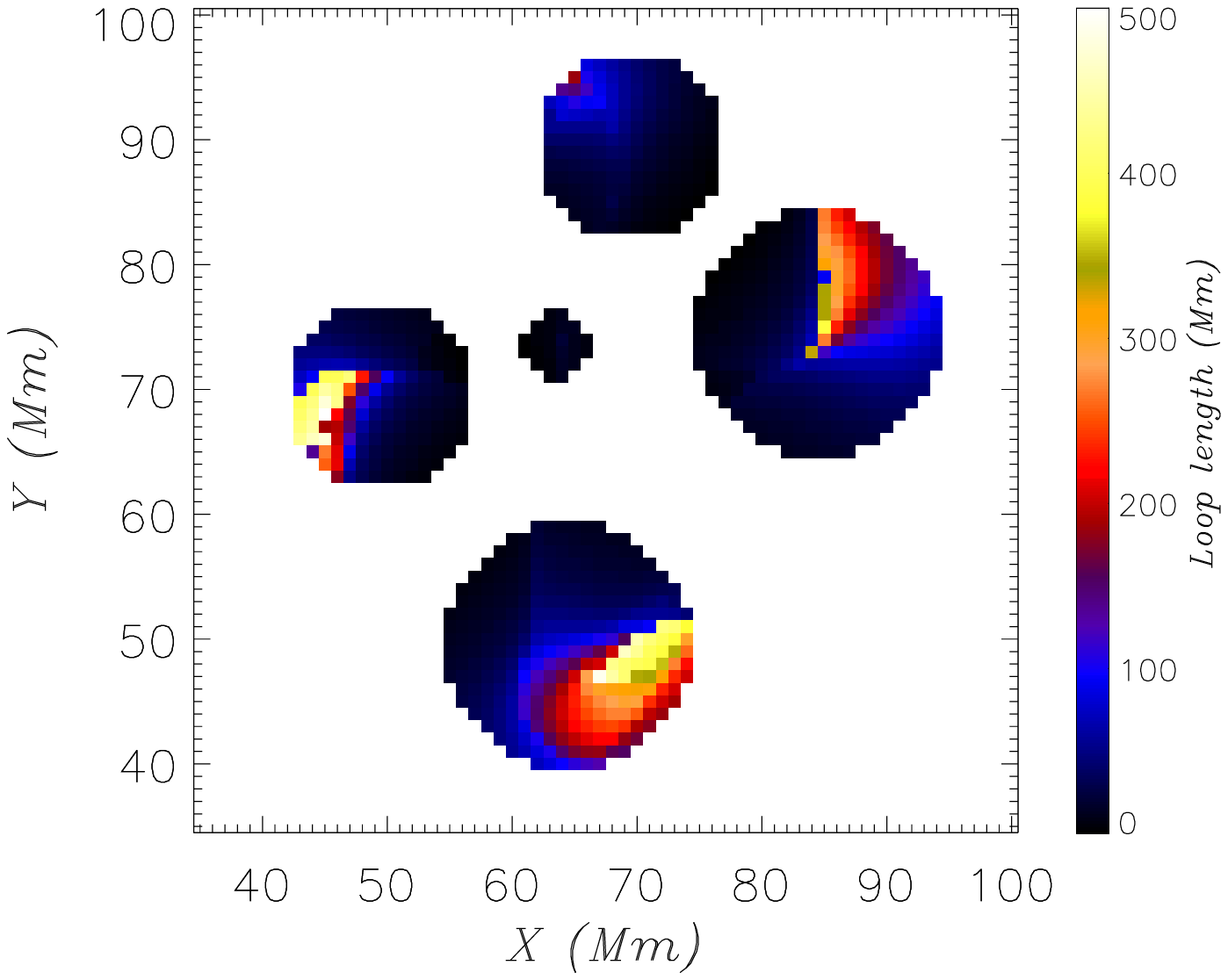}
\includegraphics[width=0.49\textwidth]{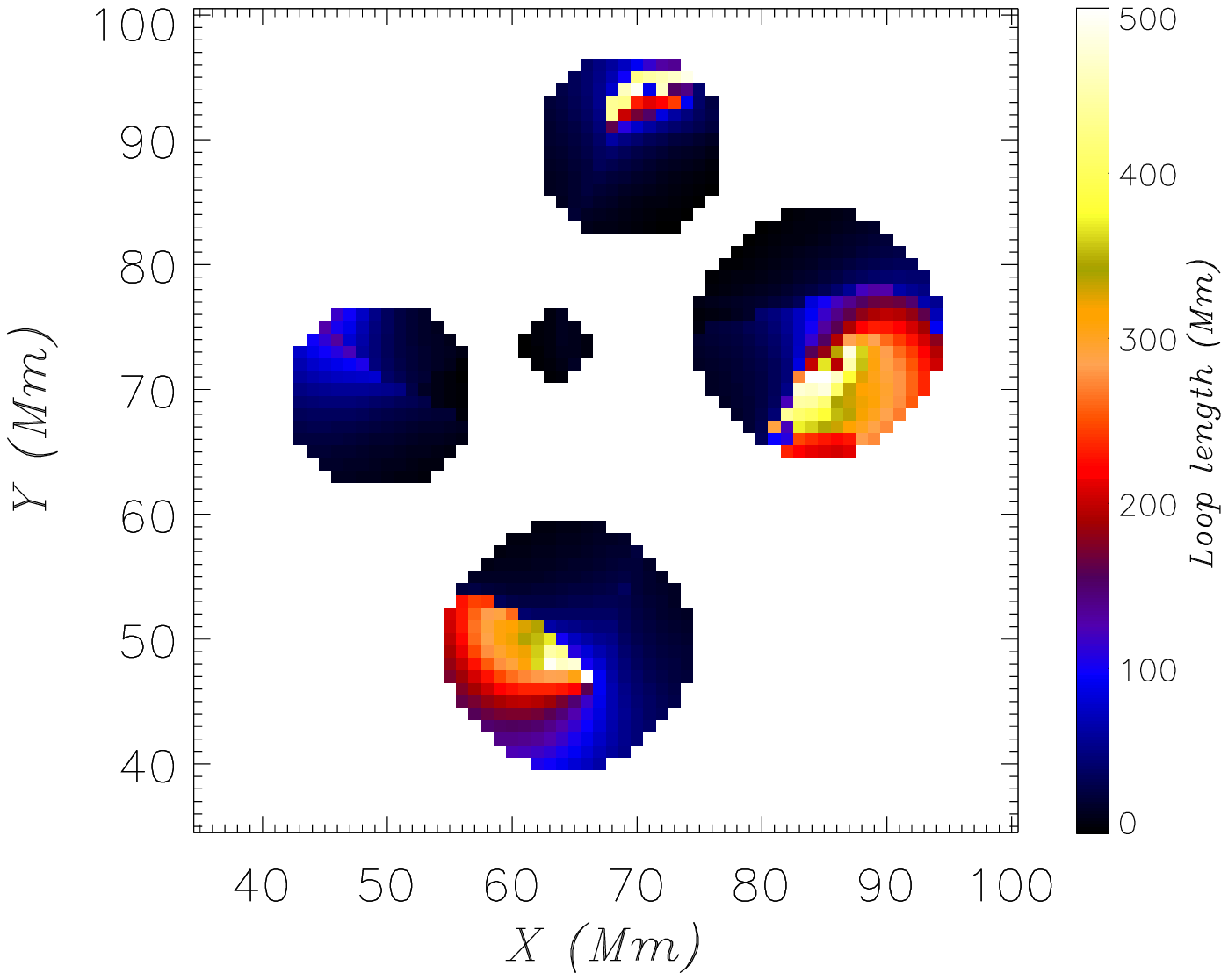}
\includegraphics[width=0.49\textwidth]{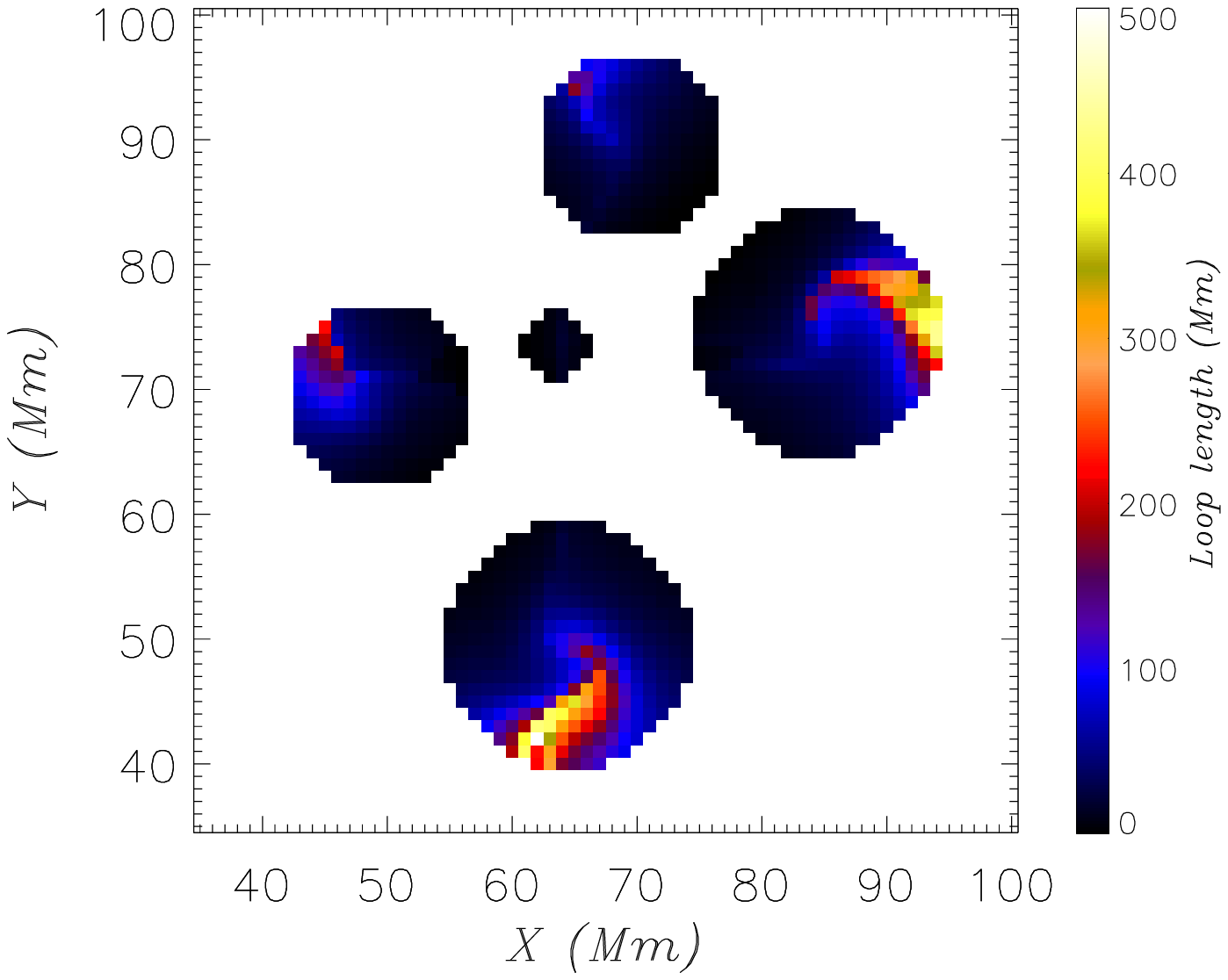}
\caption{Connectivity maps for a restricted field of view ($x$=[30, 110],
$y$=[30, 110]) depicting the field lines having the same length for the
potential field (top left), the force-free field with $\alpha = -0.62$ Mm$^{-1}$
(top right) and $\alpha = 0.74$ Mm$^{-1}$ (bottom left), and the nonlinear
force-free field  with $J_{z0} = 20$ mA m$^{-2}$ (bottom right).  The colour bar
indicates the loop lengths in Mm.}
\label{fig:connect}
\end{figure}

To depict the connectivity of the field lines, we associate, on a 2D map, the
location of the footpoint of a field line to their length. The connectivity
plots are drawn for the four characteristic force-free fields in
Figure~\ref{fig:connect}. We first notice that the connectivity of the parasitic
polarity is not modified for the different models. The changes of connectivity
mostly affect the field lines on the outer edges of the polarities and not the
regions of strong magnetic-field concentrations. It is worth noticing that the
Grad--Rubin algorithm computes the nonlinear force-free field from the
positive polarities. For the linear force-free fields, we notice that the long
field lines are moved counter-clockwise (clockwise) when the negative
(positive) values of $\alpha$ are increased in absolute value. For the
nonlinear force-free field, a positive (negative) value of $J_{z0}$ gives
the same behaviour as a negative (positive) value of $\alpha$ in the
linear force-free configurations. Therefore the connectivity of field lines
depends strongly on the amount of electric currents injected in the magnetic
configurations.  

\section{Null Point Properties} \label{sec:topo}

	\subsection{Location of Null Points}
	
\begin{figure}
\centering
\includegraphics[width=.49\textwidth]{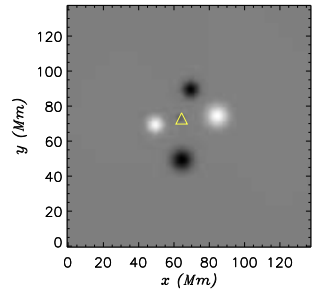}
\caption{Location of the null point (yellow triangle) located at $z$ = 6.9 Mm in
the potential field configuration. The background image is the distribution of the
vertical component of the magnetic field (black and white for negative and positive
polarities).}
\label{fig:np_pot_loc}
\end{figure}

\begin{figure}
\centering
\includegraphics[width=0.49\textwidth]{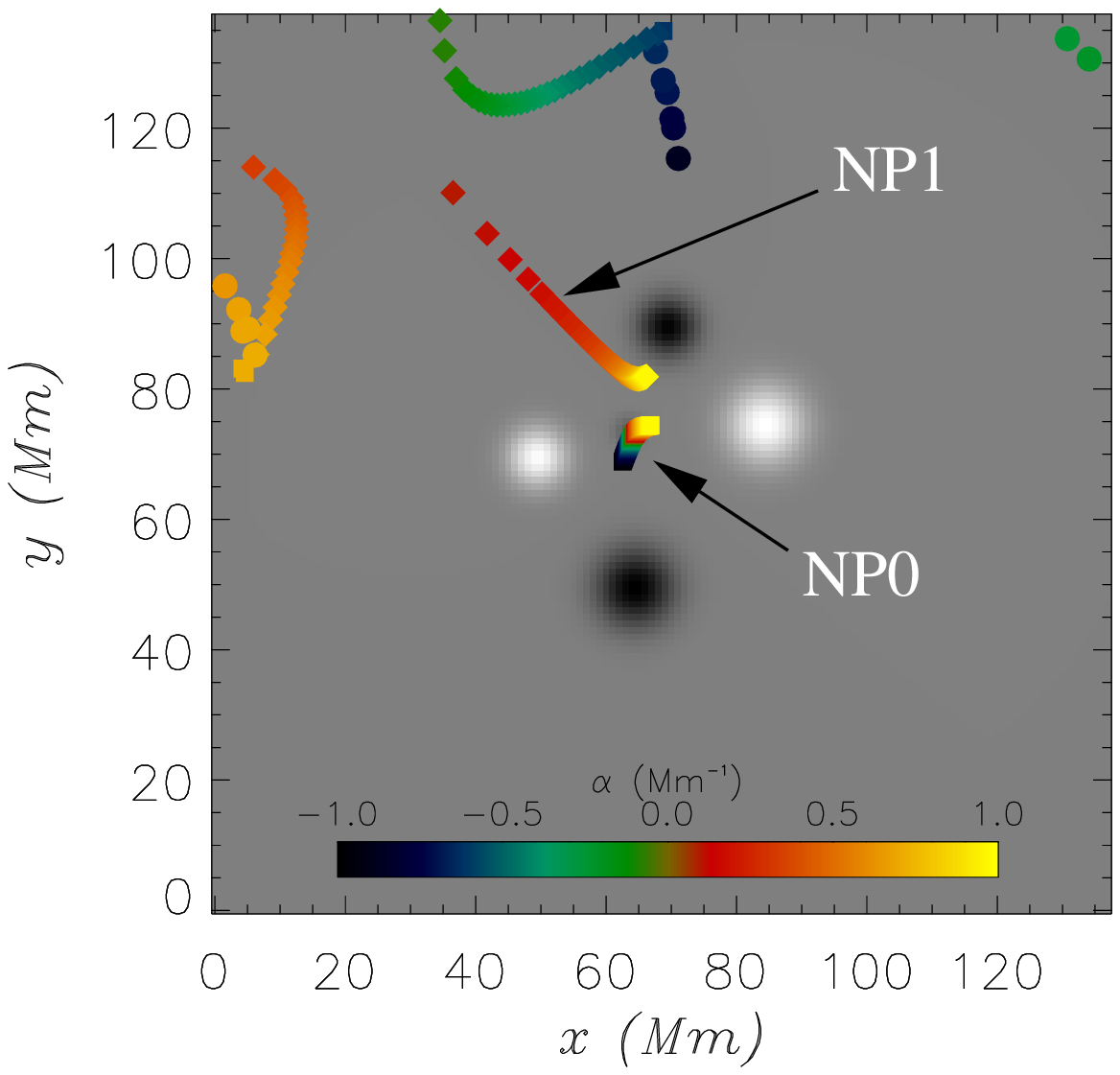}
\includegraphics[width=0.49\textwidth]{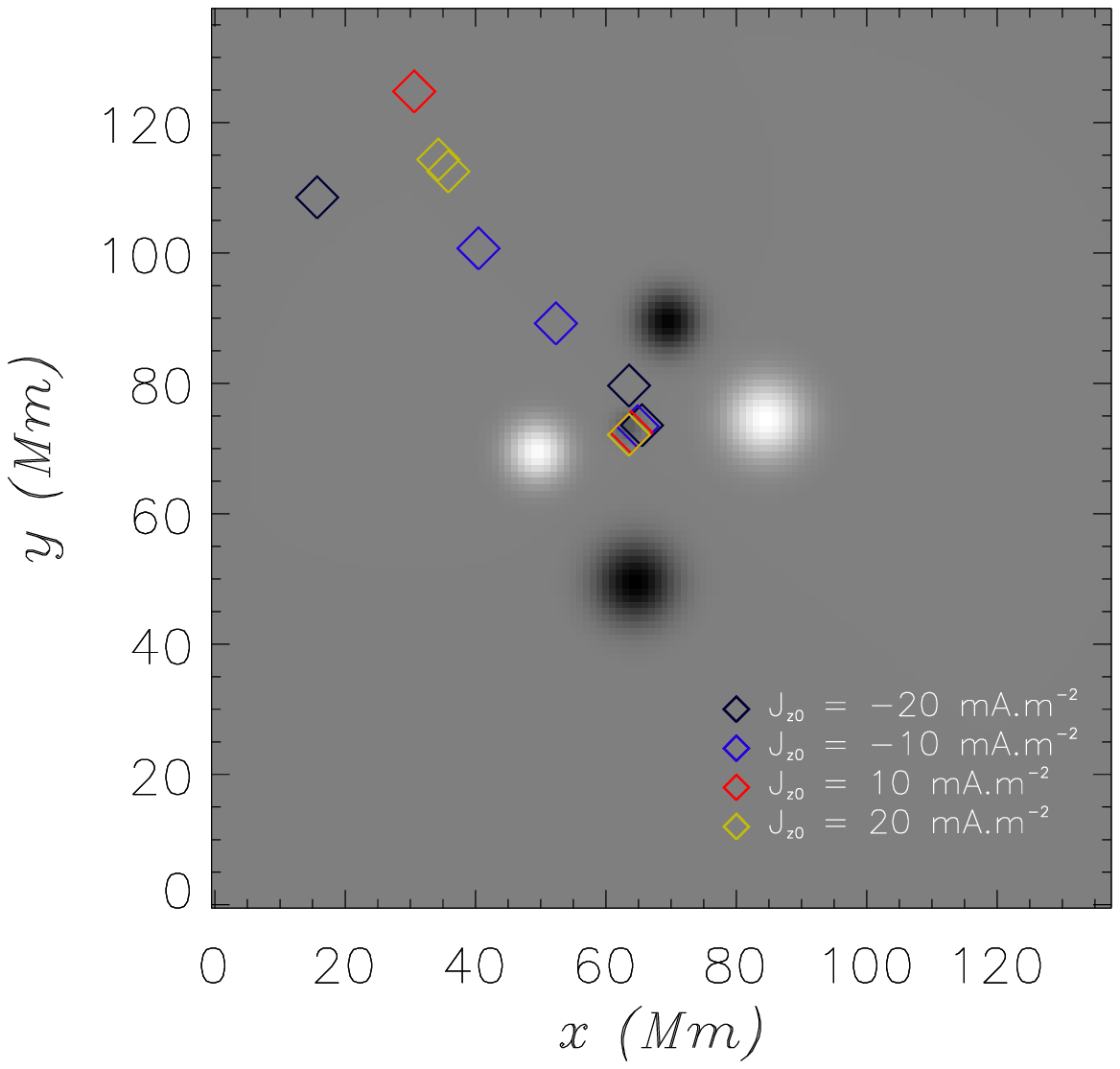}
\caption{Same as Figure~\ref{fig:np_pot_loc}: (Left) linear force-free fields
with  $\alpha$ varying from -1 to 1 Mm$^{-1}$. NP0 is depicted by a square
symbol whilst other null points are depicted by a diamond (circle) symbol
when their height is greater (lower) than NP0. (Right) nonlinear
force-free fields with $J_{z0} = [-20, -10, 10, 20]$ mA m$^{-2}$ (diamonds).}
\label{fig:np_loc}
\end{figure}

We locate the null points within the magnetic configurations using the trilinear
interpolation method developed by \inlinecite{hay07}. We then compare the location of
the null points for the different models.  We plot the location of the null
points onto the $x$--$y$-plane for the potential field in Figure~\ref{fig:np_pot_loc},
for the linear force-free fields with $\alpha$ between --1 and 1 Mm$^{-1}$ in
Figure~\ref{fig:np_loc} left, and for the nonlinear force-free fields using
several values of $J_{z0}$ in Figure~\ref{fig:np_loc} right. There is only one
null point NP0 in the potential field located 6.9 Mm above the parasitic
negative polarity (see Figure~\ref{fig:np_pot_loc}). In Figure~\ref{fig:np_loc}, the
null points present in the magnetic configuration for both the linear and
nonlinear force-free fields can be divided into two groups:
\begin{itemize}
\item[{\em i})]{near the location of the potential null point, a null point is found
for all models whatever the value of the current density or the force-free
parameter $\alpha$;}
\item[{\em ii})]{other null points can appear mostly near the boundaries but also in
strong field regions.}
\end{itemize} 
For the first group, we conclude that the null point NP0 created in the
potential field is stable in the other models and its location is just slightly
influenced by the current density: the null point is moving up and down, left
and right depending on the sign of the current density. For the second group, we
can already notice that most of the null points are located near the side
boundaries, and in addition the null point NP1 is moving towards strong field
regions when $\alpha$ is increased (see Figure~\ref{fig:np_loc} left). However, we
need to investigate the properties of the null point to draw conclusions (see
Section~\ref{sec:prop}). 

\begin{figure}
\centering
\includegraphics[width=1.\linewidth]{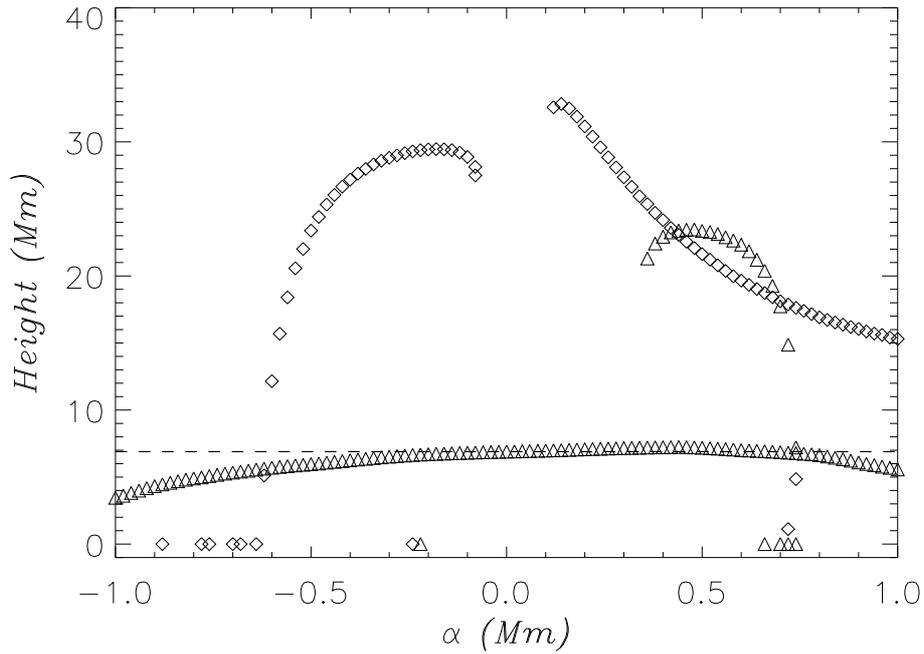}
\caption{Height of the null points in the potential and linear force-free field
models for $\alpha$ varying from --1 to 1 Mm$^{-1}$. The dashed line indicates
the height of NP0 in the potential configuration. Triangles (diamonds) indicate
positive (negative) null points.}
\label{fig:np_sign}
\end{figure}

In Figure~\ref{fig:np_sign}, we plot the distribution of null points as a
function of height for the linear force-free configurations. We indicate the
sign of the null points: triangles (resp. diamonds) for positive (resp.
negative) null points. This plot allows us to track the null points depending on
the value of the parameter $\alpha$ and on the sign of the null points. We
notice that the potential-field null point NP0 evolves smoothly when the
$\alpha$ parameter (in absolute value) increases: the null point height varies
continuously from 4 Mm to 7.5 Mm whilst the null point height in the potential
field is 6.9 Mm as indicated by the dashed line in Figure~\ref{fig:np_sign} (see
also Table~\ref{tab:prop}). 

Other null points appear when $|\alpha| > 0.1$ Mm$^{-1}$ (see
Figure~\ref{fig:np_sign}). We obtain up to five null points at $\alpha \approx $
0.7 Mm$^{-1}$ (see Table~\ref{tab:prop}). Several null points are located near
the bottom boundary. All null points are at a height less than 40 Mm (one-third
of the vertical length of the computational box) where the bipolar field is
dominant: the complexity of the quadrupolar and parasitic polarities is located
below 40 Mm. Note that the bipolarisation of the magnetic field and its
associated height are also measures of the complexity of the magnetic field. 

The height of the stable null point decreases when $\alpha$ is negative and it
increases when $\alpha$ is positive. Nevertheless for $\alpha > 0.5$ Mm$^{-1}$,
the height starts to decrease, influenced by the null point NP1 propagating in
the strong field region towards NP0 (see Figure~\ref{fig:np_loc} left). 

In Figure~\ref{fig:np_loc} right, we find the same groups of null points for the
nonlinear force-free fields. We note that, for this experiment, a nonlinear
force-free field with negative (positive) values of $J_{z0}$ has the same
behaviour as a linear force-free field with positive (negative) values of
$\alpha$.

	\subsection{Properties} \label{sec:prop}
	
As discussed in Appendix A, the null points can be classified as
negative and positive depending on whether the fan field lines are radiating in or out,
respectively, from the null point. To derive the spectral properties of a null
point, we first need to derive the Jacobian matrix and then to compute the
eigenvalues. To describe these properties, we thus introduce a quantity
which helps us to classify the nature of the null points:
the spectral radius of the Jacobian matrix ($\rho_{J}$) as follows: 
\begin{equation}
\rho_{J} = \max_{i}{(|\lambda_i|)} \quad \mathrm{for} \quad i=1,2,3
\end{equation}
where $\lambda_{i}$ are the (real or complex) eigenvalues of the Jacobian matrix.
For a Jacobian matrix having two complex-conjugate eigenvalues, the spectral
radius ($\rho_J$) is the real eigenvalue for a divergence-free magnetic field.

In Table~\ref{tab:prop}, we summarize the properties of the null points (type,
location, eigenvalues, spectral radius) for the potential field, the linear
force-free fields with $\alpha = [-0.62, 0.74]$ Mm$^{-1}$, and the nonlinear
force-free fields with $J_{z0} = [-20, -10, 10, 20]$ mA m$^{-2}$. The
eigenvalues are sorted in such way that the spectral radius corresponds to the
absolute value of the first eigenvalue. The null point present in the potential
configuration is a positive null point with one negative and two positive
eigenvalues. This null point (NP0) is also present in the linear and nonlinear
force-free configurations with a slight displacement up and down, left and right
depending on the sign of the $\alpha$ values. Thus, the null point (NP0)
originally created in the potential field is stable for all force-free models.
In particular, NP0 is always a positive null point and the spectral radius is
almost constant. Therefore we conjecture that the spectral radius gives a good
proxy for the stability of a null point in a magnetic-field configuration. Note
that a large spectral radius indicates large magnetic-field gradients.

In addition to NP0, other null points can appear in the magnetic configuration
depending on the strength of the current density. It is noticeable that null
points are mostly created in pairs (negative and positive null points) or they
have complex eigenvalues. The latter case which indeed cannot exist in the
force-free assumption corresponds to null points appearing at locations where
{\em i}) the Taylor expansion in the vicinity of the null point is not valid
anymore, or {\em ii}) the null-point finder algorithm breaks down, or {\em iii})
the Jacobian matrix elements cannot be derived with enough accuracy (especially
in weak-field regions). Note that we only found four values of $\alpha$ (among 120)
for which complex conjugate eigenvalues exist.  

\begin{table}
\caption{Properties of null points (type, location, eigenvalues) for the
potential, linear  force-free fields with $\alpha = [-0.62, 0.74]$ Mm$^{-1}$ and
nonlinear force-free fields with $J_{z0} = [-20, -10, 10, 20]$ mA m$^{-2}$. $^a$
One real and two complex eigenvalues (only the real parts are reported).}
\label{tab:prop}
\begin{tabular}{cccccc}
\hline \\[-0.2cm]
Model & $\alpha$ or $J_{z0}$  & Type & Location & Eigenvalues \\
 & & & ($x_0$, $y_0$, $z_0$) & ($\lambda_1 = \pm \rho_J$, $\lambda_2$,
 	$\lambda_3$)  \\
\hline \\[-0.2cm]
{\em Potential} & & + & (64.30, 72.94, 6.90) & 
	({\bf--0.026}, 0.021, 5.5 10$^{-3}$) \\[0.2cm]

\hline \\[-0.2cm]
{\em Linear}  & -0.62  & + &  (63.43, 71.2, 5.63) & 
	 ({\bf--0.032}, 0.023, 9.28 10$^{-3}$) \\
 {\em Force-Free} &  & -- & (68.8, 134.9, 5.12) & 
	(9.26, --4.8, --4.8)$\times$ 10$^{-5}$ $^{a}$ \\[0.2cm]

($\alpha$) & 0.74 & + & (65.8, 74.2, 6.78) & 
	 ({\bf--0.026}, 0.021, 5.7 10$^{-3}$) \\
 &  & + & (6.14, 85.2, 6 10$^{-4}$) & 
	(--2.68, 2.28, 0.836)$\times$ 10$^{-4}$ \\
 & & -- & (64.31, 81.63, 17.62) & 
 	(5.8, --4.5, --1.0)$\times$ 10$^{-3}$ \\
 & & -- & (4.22, 83.12, 4.85) & 
	(5.83, --3.17, --3.17)$\times$ 10$^{-5}$ $^{a}$ \\
 & & + & (4.58, 82.44, 7.21) & 
	(-10, 4.86, 4.86)$\times$ 10$^{-5}$ $^{a}$
	\\[0.2cm]

\hline \\[-0.2cm]
{\em Nonlinear} & -20 & + &  (65.54, 73.57, 6.31) & 
	({\bf--0.024}, 0.019, 4.86 10$^{-3}$) \\
 {\em Force-Free} &  & -- &  (63.54, 79.64, 17.23) & 
	(4.71, --3.78, --0.88)$\times$ 10$^{-3}$ \\
($J_{z0}$) & & + &  (15.70, 108.5, 36.98) & 
 	(--4.38, 3.72, 0.35)$\times$ 10$^{-5}$ \\[0.2cm]
 & -10 & + &  (64.84, 73.39, 6.77) & 
	 ({\bf--0.021}, 0.017, 4.55 10$^{-3}$) \\
 &  & -- & (52.33, 89.19, 30.59) & 
	(4.02, --3.83, --0.21)$\times$ 10$^{-4}$ \\
 & & + & (40.44, 100.7, 36.39) & 
 	 (--1.17, 1.13, 0.028)$\times$ 10$^{-4}$ \\[0.2cm]
 & 10 & + &  (63.87, 72.49, 7.12) & 
	 ({\bf--0.020}, 0.016, 4.24 10$^{-3}$) \\
 &  & -- & (30.59, 124.7, 0.06) & 
	(7.27, --6.62, --0.69)$\times$ 10$^{-5}$ \\[0.2cm]
 & 20 & + & (63.53, 72.10, 7.65) & 
	({\bf--0.024}, 0.019, 4.9 10$^{-3}$) \\
 &  & -- & (34.29, 114.3, 0.025) & 
	(7.6,--5.9,--1.9)$\times$ 10$^{-4}$ \\
 & & -- & (35.82, 112.5, 14.29) & 
 	(6.8, --6.3, --0.66)$\times$ 10$^{-4}$ \\
\hline
\end{tabular}

\end{table}

\section{Magnetic Energy Budget} \label{sec:nrj}
	
	\subsection{Total Magnetic Energy}

\begin{figure}
\includegraphics[width=1.\linewidth]{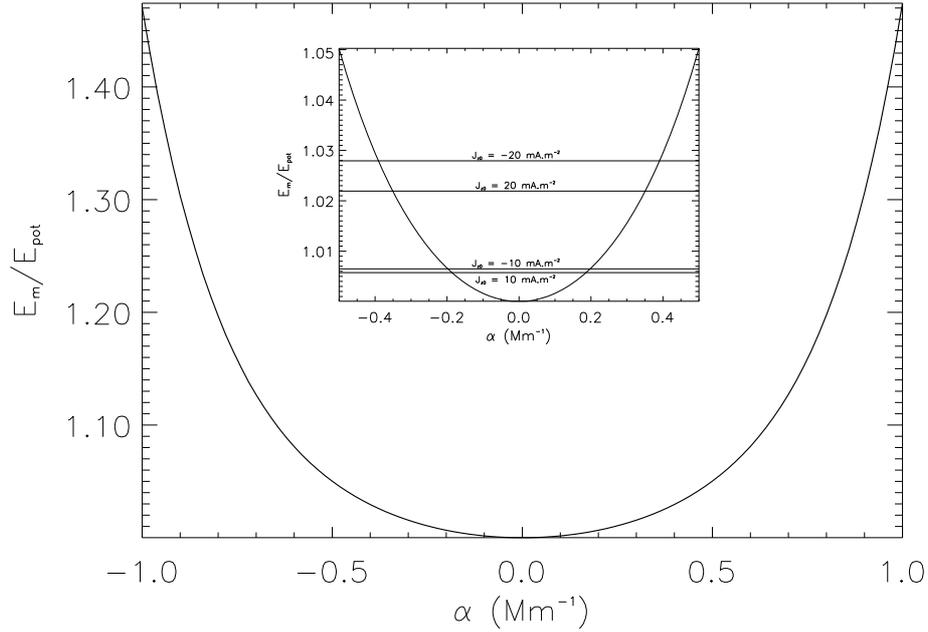}
\caption{Magnetic energy ($E_m$) relative to the potential field energy
($E_{\textrm{\tiny{pot}}}$) as a function of the force-free parameter ($\alpha$:
Mm$^{-1}$) for the linear force-free extrapolations. The solid lines indicate
the relative energy levels $E_m/E_{\textrm{\tiny{pot}}}$ for the nonlinear
force-free extrapolations for $J_{z0} = [-20, -10, 10, 20]$ mA m$^{-2}$. }
\label{fig:nrj}
\end{figure}

In Figure~\ref{fig:nrj}, we plot the magnetic energy ($E_{m}$) above the
potential-field energy ($E_{\textrm{\tiny{pot}}}$) in the computational volume
($\Omega$). As the potential field is a minimum-energy state, $E_m$ is always
above $E_{\textrm{\tiny{pot}}}$. The latter inequality is true if and only if
both the force-free and potential fields are computed with the same normal
component of the magnetic field on each side of the computational box. 

The energy curve as a function of $\alpha$ is similar to the second-order
polynomial curve obtained in Figure~11 of \inlinecite{reg07} for a solar active
region. For $\alpha$ ranging from -1 to 1 Mm$^{-1}$, the magnetic energy of
linear force-free fields is not more 50\% of the potential-field energy. The
curve of the free magnetic energy as a function of $\alpha$ is not symmetric
with respect to the potential field ($\alpha = 0$): the magnetic energy is
increasing more rapidly for the positive values of $\alpha$. 

The different levels of magnetic energy for the nonlinear force-free fields are
plotted in Figure~\ref{fig:nrj} as straight solid lines:
$E_m/E_{\textrm{\tiny{pot}}} = [1.0279, 1.0065, 1.0058, 1.022]$ for $ J_{z0} =
[-20, -10, 10, 20]$ mA m$^{-2}$  respectively. This small amount of magnetic
energy stored in the nonlinear force-free configurations (less than 3\%)  is a
consequence of the particular current distribution: the ring distribution has no
net current in a single polarity and return currents on the edges of the flux
bundles which confine the magnetic field in strong-field regions without
generating twisted flux bundles.  
	
	\subsection{Energy Density Distributions}

\begin{figure}
\centering
\includegraphics[width=0.49\textwidth]{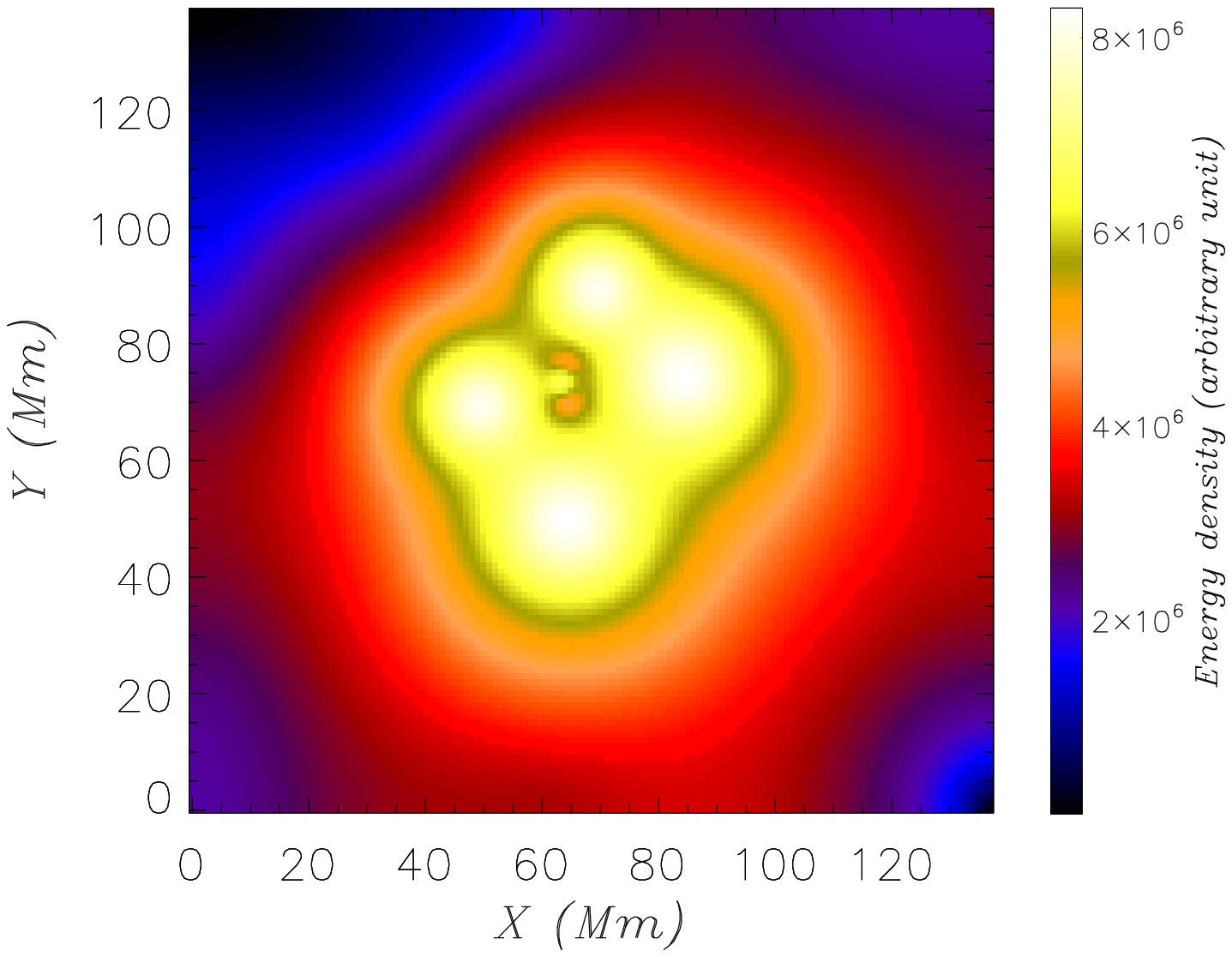}
\includegraphics[width=0.49\textwidth]{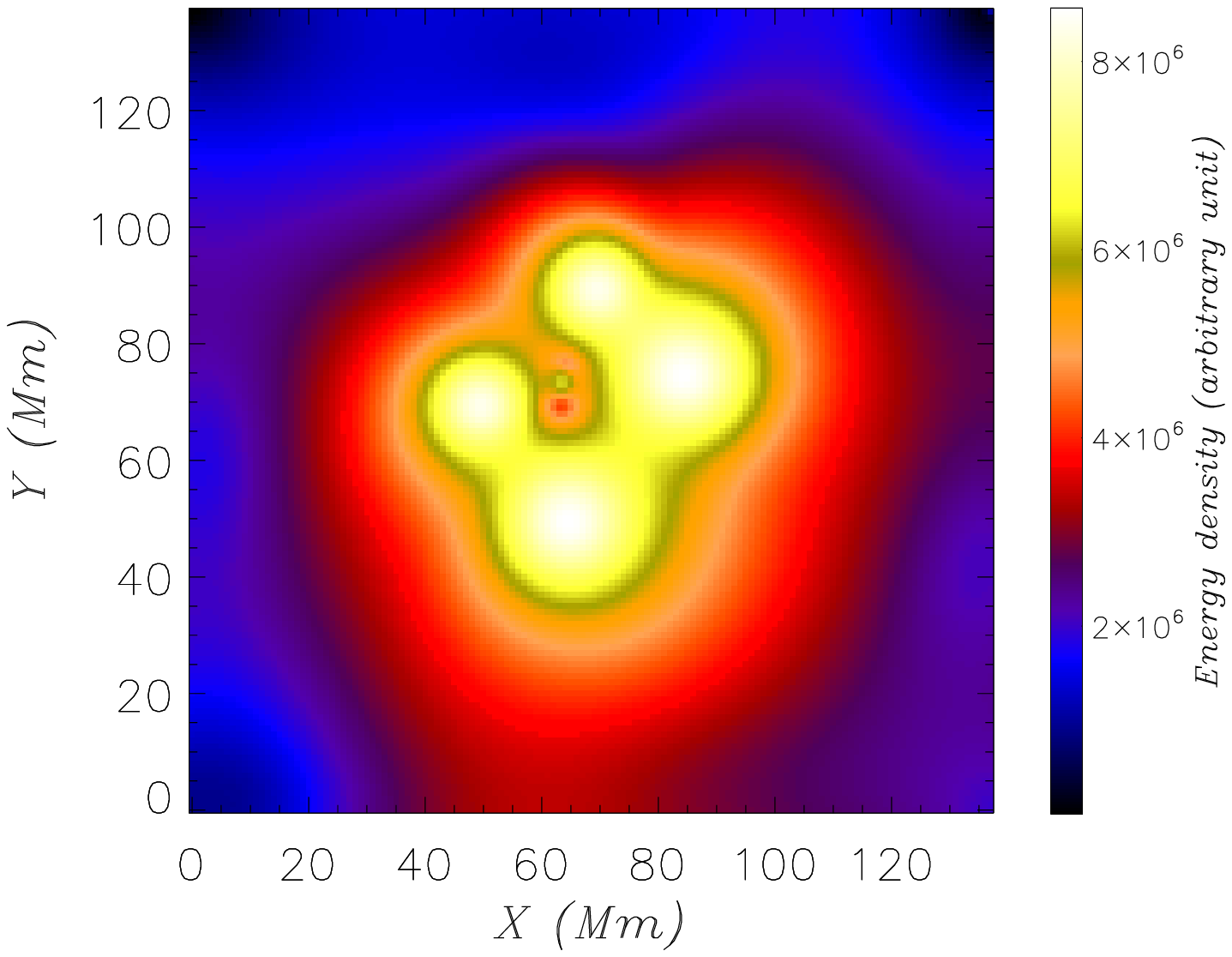}
\includegraphics[width=0.49\textwidth]{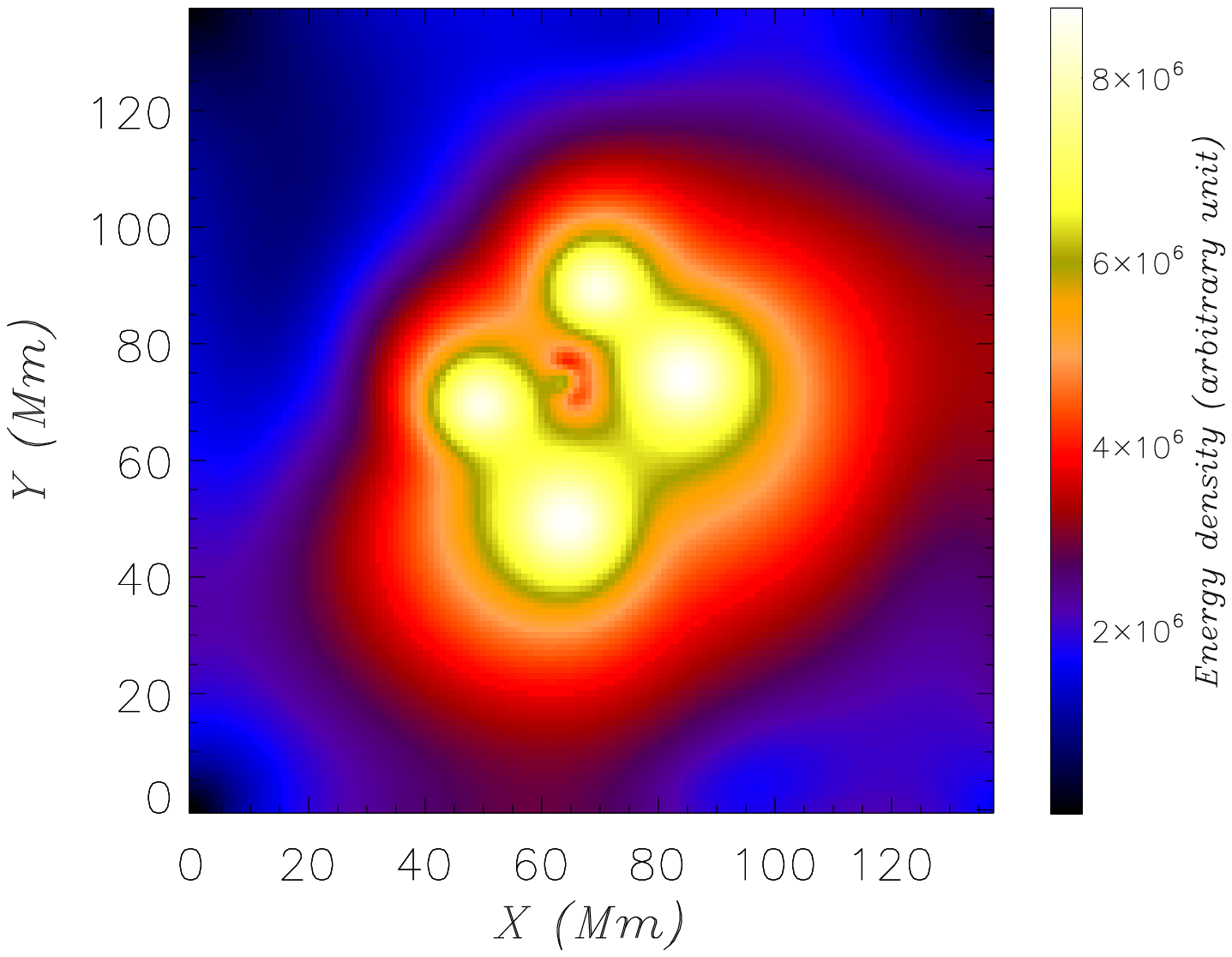}
\includegraphics[width=0.49\textwidth]{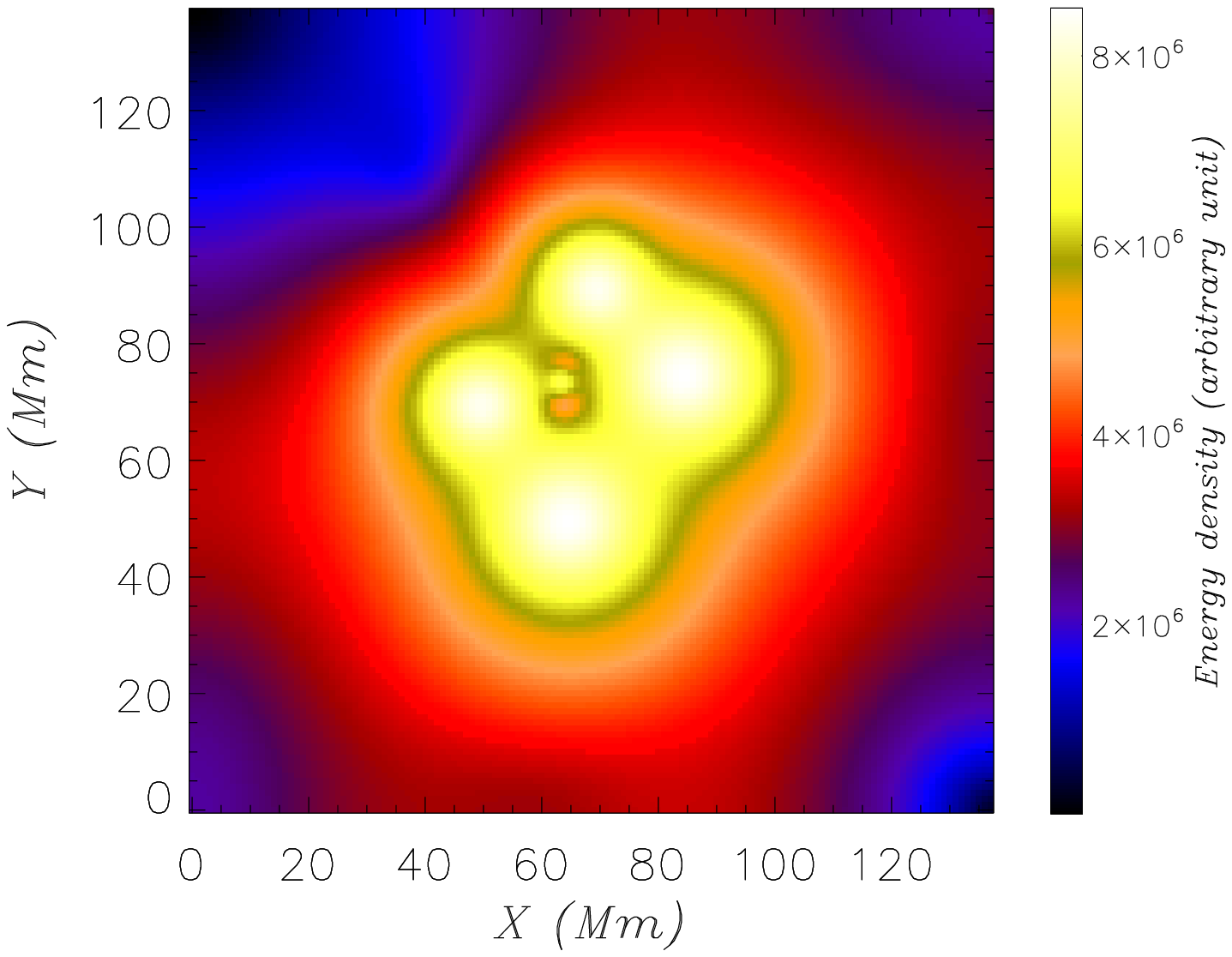}
\caption{Energy-density maps integrated along the $z$-axis (in arbitrary unit)
for the potential field (top left), the force-free field with $\alpha = -0.62$
Mm$^{-1}$ (top right) and $\alpha = 0.74$ Mm$^{-1}$ (bottom left), and the
nonlinear force-free field  with $J_{z0} = 20$ mA m$^{-2}$ (bottom
right).  The colour bar indicates the energy density in arbitrary units.}
\label{fig:nrj_den}
\end{figure}

In Figure~\ref{fig:nrj_den}, we plot the magnetic-energy density integrated
along the $z$-axis to study the distribution of the magnetic energy of the four
magnetic fields analysed above. For the potential field, the distribution of the
energy density is dominated by the energy (or field strength) near the bottom
boundary for the five magnetic polarities. We notice that there is a local
minimum of magnetic-energy density where the null point NP0 is located. In
addition, there is another obvious local minimum located on the other side of
the parasitic polarity with respect to NP0 (location: $x = 62$, $y = 78$). For
the two linear force-free fields, the magnetic-energy density distribution is
dominated by the four polarities of the initial quadrupolar field where the
magnetic-field strength is large, whilst the parasitic polarity does not
influence the distribution. Again we notice that there is a strong local minimum
at the location of NP0 and in addition there is an annulus-like series of local
minima connecting NP0 and NP1 around the parasitic polarity. The other null
points located in weak-field regions are not observed on the energy-density
maps. For the nonlinear force-free field, the energy density distribution looks
very much like the distribution of the potential field with a maximum of energy
density slightly increased.   

We notice that we are able to easily identify null points as local minima in
the distribution of energy density where strong magnetic-field gradients are
observed (large spectral radius).

\section{Electric Currents} \label{sec:cur}

\begin{figure}
\centering
\includegraphics[width=0.32\textwidth]{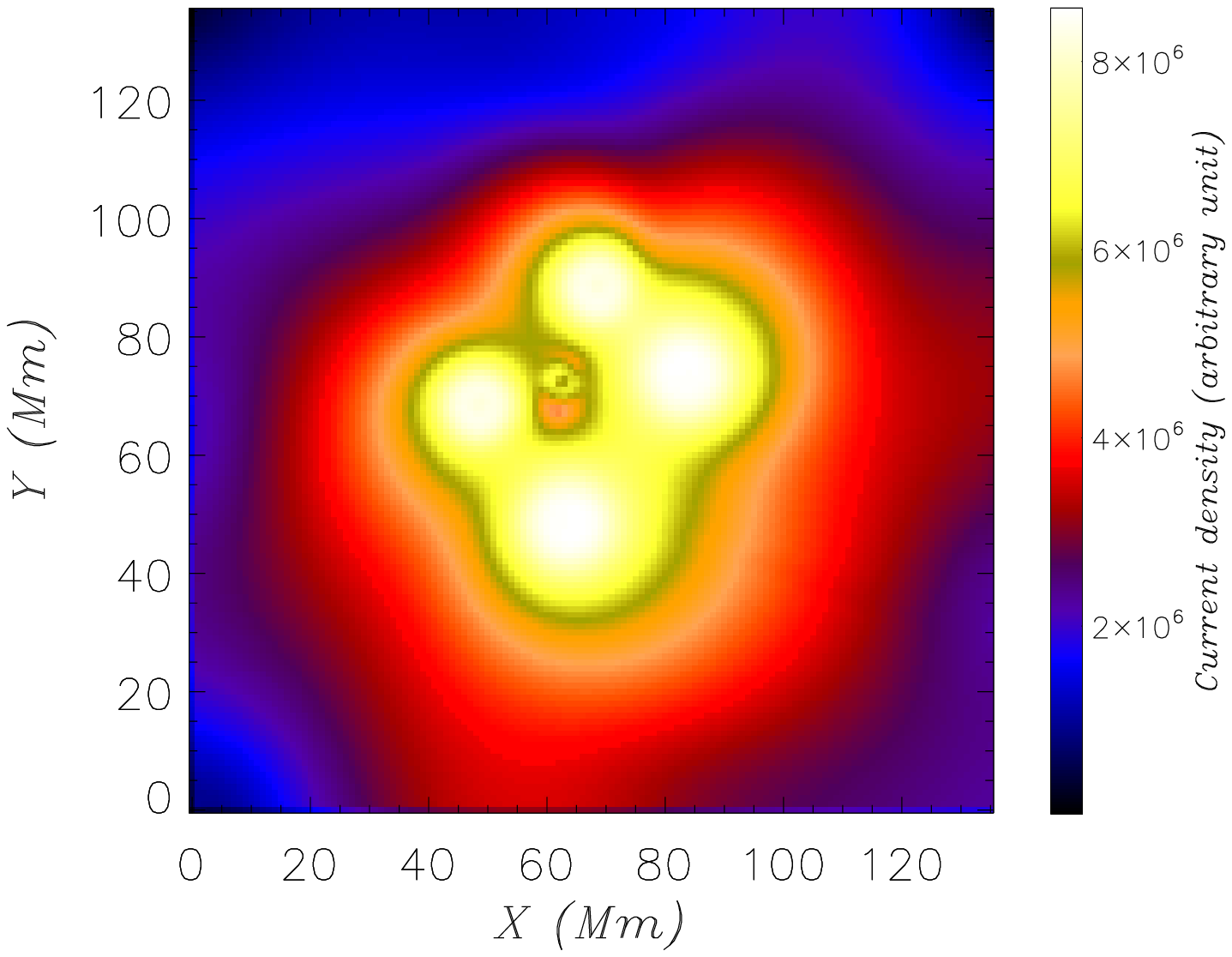}
\includegraphics[width=0.32\textwidth]{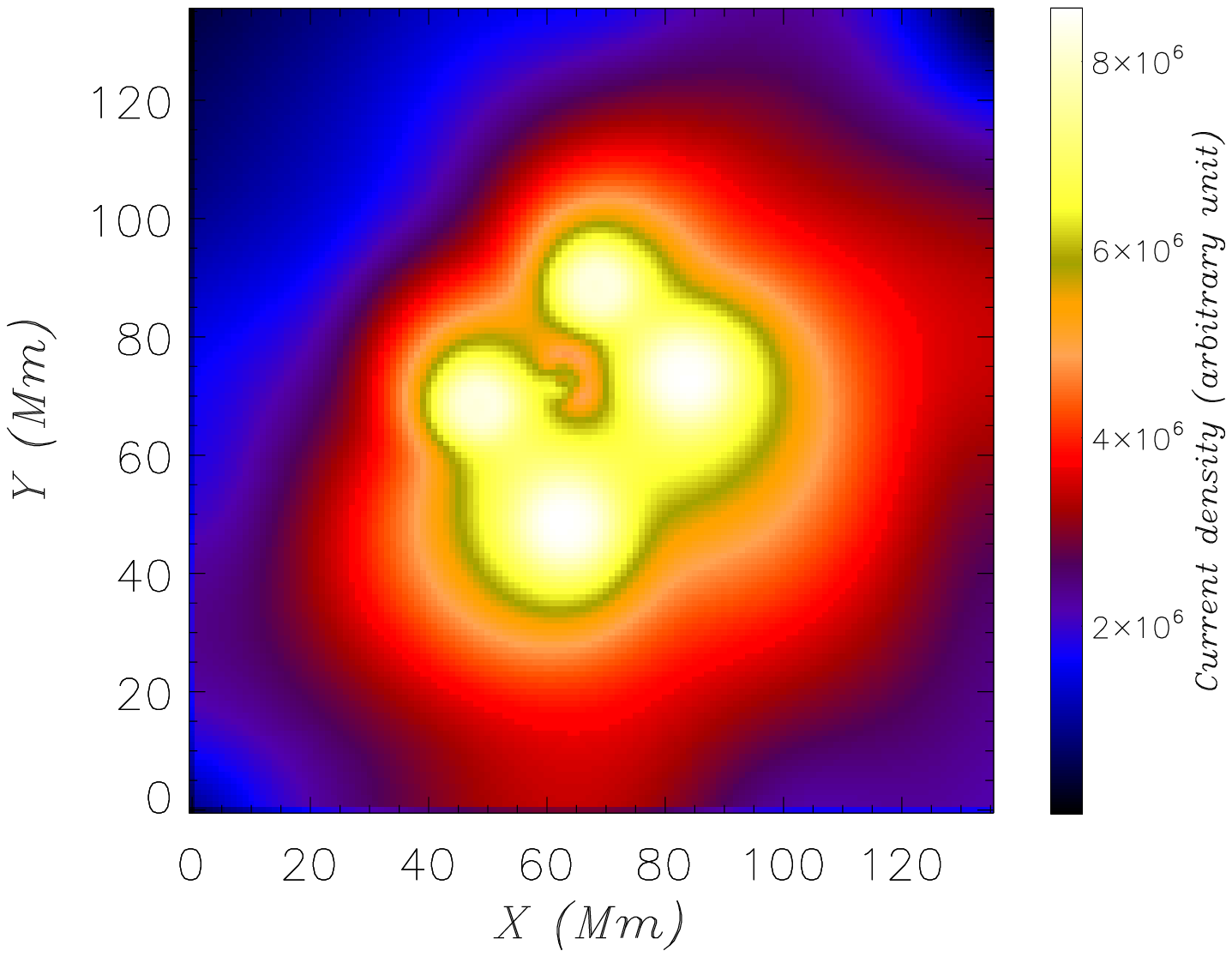}
\includegraphics[width=0.32\textwidth]{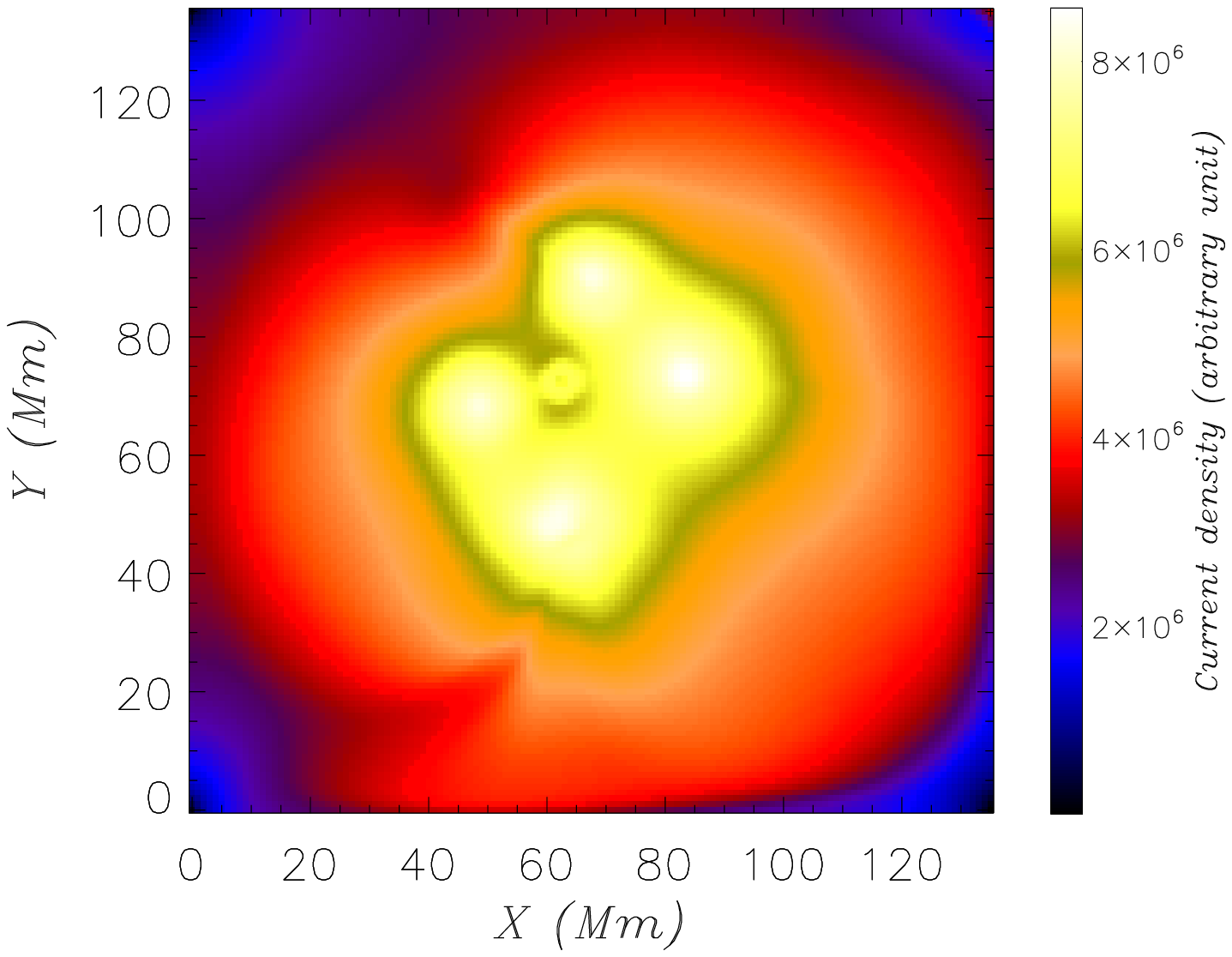}
\caption{Electric-current density maps integrated along the $z$-axis for the
force-free field with $\alpha = -0.62$ Mm$^{-1}$ (left) and $\alpha = 0.74$
Mm$^{-1}$ (middle), and the nonlinear force-free field  with $J_{z0} = 20$
mA m$^{-2}$ (right).  The colour bar indicates the current density
in arbitrary unit.}
\label{fig:cur_den}
\end{figure}

In Figure~\ref{fig:cur_den}, we plot the electric-current density integrated
along the $z$-axis for three of the four characteristic magnetic field
computations. We have computed the three components of the current density from
the curl of the magnetic field and then plotted the current-density strength (or
modulus). The potential field has zero electric current (or only tiny currents
due to the errors when the magnetic-field components are differentiated). The
distribution of the current density is different from the distributions of the
energy density (see Figure~\ref{fig:nrj_den}). The five polarities contribute
a large amount to the current-density distribution. As noticed for the
energy-density distribution, there exists a local minimum where the null point
NP0 is located, and an annulus-like series of local minima exists connecting the
null points NP0 and NP1 in the linear force-free fields. For the nonlinear
force-free field configuration, the electric-current density distribution is
similar to the linear force-free distribution with a local minimum at the
location of NP0. 

Except for the location of null points, it is not obvious wwhere to locate with
this method the other topological elements where the current density is supposed
to be increased. This shows that, for this configuration, the storage of current
density along topological elements is not an efficient mechanism compared to
the current density stored in the strong-field regions above the magnetic 
polarities. 

\section{Discussion and Conclusions} \label{sec:concl}

We investigated the changes in the magnetic-field configurations obtained for
different force-free models (potential, linear, and nonlinear force-free fields)
using the same boundary conditions. We analysed the changes in terms of geometry
and connectivity of field lines, magnetic energy, and electric-current
distributions. We performed this analysis for a continuous magnetic-field
distribution with no symmetry. We imposed an electric-current distribution that
we have proven to be stable when a large amount of current is injected into the
magnetic configuration \cite{reg09a}. Despite the previous works on this topic,
we have here provided an extended analysis of magnetic configurations with a
topology that has never been performed before. 

For this experiment, the initial configuration corresponds to a potential field
with five sources (two large bipoles and one parasitic polarity) having a null
point NP0 in the corona. By injecting currents in the magnetic configuration,
the geometry and topology of the linear and nonlinear force-free configurations
are modified such that:

\begin{itemize} 
\item{the geometry of the field lines is modified similarly to previous
results obtained on the same topic: statistically, the field lines are higher
and longer when the absolute value of electric current is increased;}
\item{the connectivity of the field lines can be strongly modified near the
topological elements (where the connectivity is changed rapidly);}
\item{the initial null point NP0 is moved slightly up or down when the
force-free parameter ($\alpha$) varies but remains with the same basic
properties, in particular the spectral radius remains almost constant;}
\item{other null points (up to five) can appear in the magnetic configurations;
most of them are located near the boundaries but one (NP1) which propagates
towards the strong-field region when the current density is increased;}
\item{the magnetic energy and current distributions can highlight the location
of stable null points where strong magnetic-field gradients are present.}
\end{itemize}   

We also noticed that for reasonable values of the electric current injected or
the force-free parameter $\alpha$ ($<$ 0.25 Mm$^{-1}$ in this experiment) the
magnetic configuration is almost not modified compared to the potential-field
configuration.

We thus state that null points existing in potential-field configurations are
also present in force-free configurations with the same properties ({\em e.g.}, sign,
spectral radius). This statement means that the null points found in a potential
field with a large spectral radius can be considered as stable null points in
other magnetic-field models. Even if true for this experiment we need to confirm
this statement in a future statistical study of solar active region magnetic
fields. It is important to note here that this statement is true {\em i}) when the
force-free fields are computed with the same boundary conditions, {\em ii}) when
there is no noise in the datasets. In Appendix~\ref{sec:appb}, we noe that
the null points in potential-field configurations are slightly
affected by boundary conditions (periodic, closed, ...), spatial resolution or
size of field-of-view: the null point with the strongest spectral radius remains
stable.  

From this study, we also provide a benchmark for analysing the topology of a
magnetic configuration: it is possible to retrieve important information about
the topology just by analysing the distribution of magnetic energy density and
of the electric current density in the volume. Moreover, we emphasize the
importance of
checking the divergence-free property of the magnetic field in the vicinity of
a null point.  

\begin{acks}
SR thanks Eric Priest and Claire Parnell (University of St Andrews) for
fruitful discussions on this topic. The computations of force-free field
extrapolations have been performed using the {\sf XTRAPOL} code developed by T. Amari
(Ecole Polytechnique, France).
\end{acks}

\bibliographystyle{spr-mp-sola}



\appendix

\section{Null Point Description}

As a first approximation, we assume that the magnetic field around the null
point approaches zero linearly. The magnetic field ($\vec B$) near a neutral point
can then be expressed as a first order Taylor expansion: 
\begin{equation} 
\vec B = M \cdot \vec r, 
\end{equation} 
where $M$ is the Jacobian matrix with elements $M_{ij} = \partial B_i / \partial
x_j$ for all $i,j = 1,2,3$ and $\vec r$ is the position vector $(x, y, z)$. The
Jacobian matrix has some interesting properties: 
\begin{itemize}
\item[{\em i})]{as the magnetic field is divergence-free, $\textrm{Tr}(M) = 0$. This
property holds at each point of the magnetic field configuration.} 
\item[{\em ii})]{for potential and force-free fields which satisfy $\vec \nabla
\wedge \vec B = \vec 0$ or $\vec \nabla \wedge \vec B =\alpha \vec B$, $M$ is
symmetric at the location of the null point.  Therefore, $M$ has three real
eigenvalues and the eigenvectors are orthogonal. }
\end{itemize}
From the first property and for force-free fields, we obtain the following
relationship between the three real eigenvalues ($\lambda_1, \lambda_2,
\lambda_3$):
\begin{equation}
\lambda_1 + \lambda_2 + \lambda_3 = 0,
\end{equation}
meaning that two eigenvalues, $\lambda_2$ and $\lambda_3$ say, have the same
sign, and that $|\lambda_1| = |\lambda_2 + \lambda_3|$ implying that
$|\lambda_1| > |\lambda_2|,~|\lambda_3|$. According to \inlinecite{par96a}, the
single eigenvalue ($\lambda_1$) defines the direction of the spine field line
whilst the two other eigenvalues ($\lambda_2$, $\lambda_3$) indicate the
directions of the fan plane. Therefore we can classify null points as positive
null points with two positive and one negative eigenvalues, and negative null
points with two negative and one positive eigenvalues. For a positive (negative)
null point, the magnetic field lines in the fan plane are directed away from
(towards) the null point, whereas the spine field line is pointing towards (away
from) the null point. Following \inlinecite{hor96}, a null point is unstable if
$\det{(M)} = 0$. The determinant of $M$ is
\begin{equation}
\det{(M)} = \prod_{i=1}^{3} \lambda_i,
\end{equation}
with $\det{(M)} < 0$ for a positive null point and $\det{(M)} > 0$ for a negative
one. For potential and force-free fields, null points are stable, except if one
eigenvalue ($\lambda_2$ or $\lambda_3$) vanishes reducing the null point to a 2D
null point unstable in the 3D configuration. If $\lambda_1$ vanishes, then all
eigenvalues have to be zero, and thus the first-order Taylor expansion is no
longer valid and the null-point properties are thus derived from the Hessian
matrix instead of the Jacobian matrix.  

\section{On the Topology of Potential Fields} \label{sec:appb}

In this article, we have studied in detail the topology of force-free fields for
a distribution of photospheric magnetic field inducing the existence of a null
point. In order to compare the magnetic energy and properties of the different
configurations, we have used the same boundary conditions. But is a magnetic
configuration modified when the boundary condition are changed? and is the
topology influenced by the spatial scales of the bounded box? To address these
question, we perform a comparison of different potential fields and different
spatial scales.

\subsection{Effect of Boundary Conditions}

\begin{figure}[!ht]
\centering
\includegraphics[width=.49\textwidth]{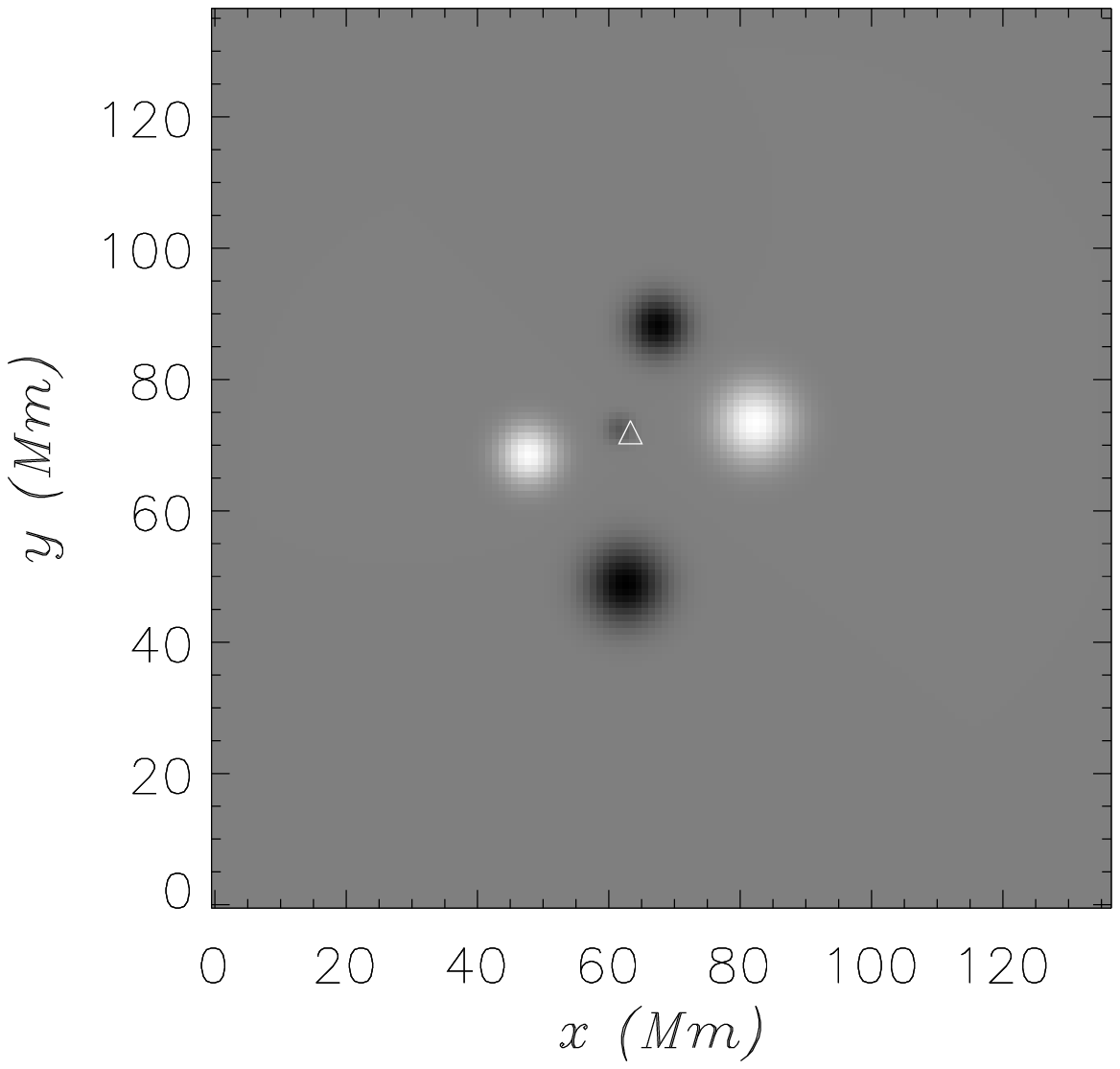}
\includegraphics[width=.49\textwidth]{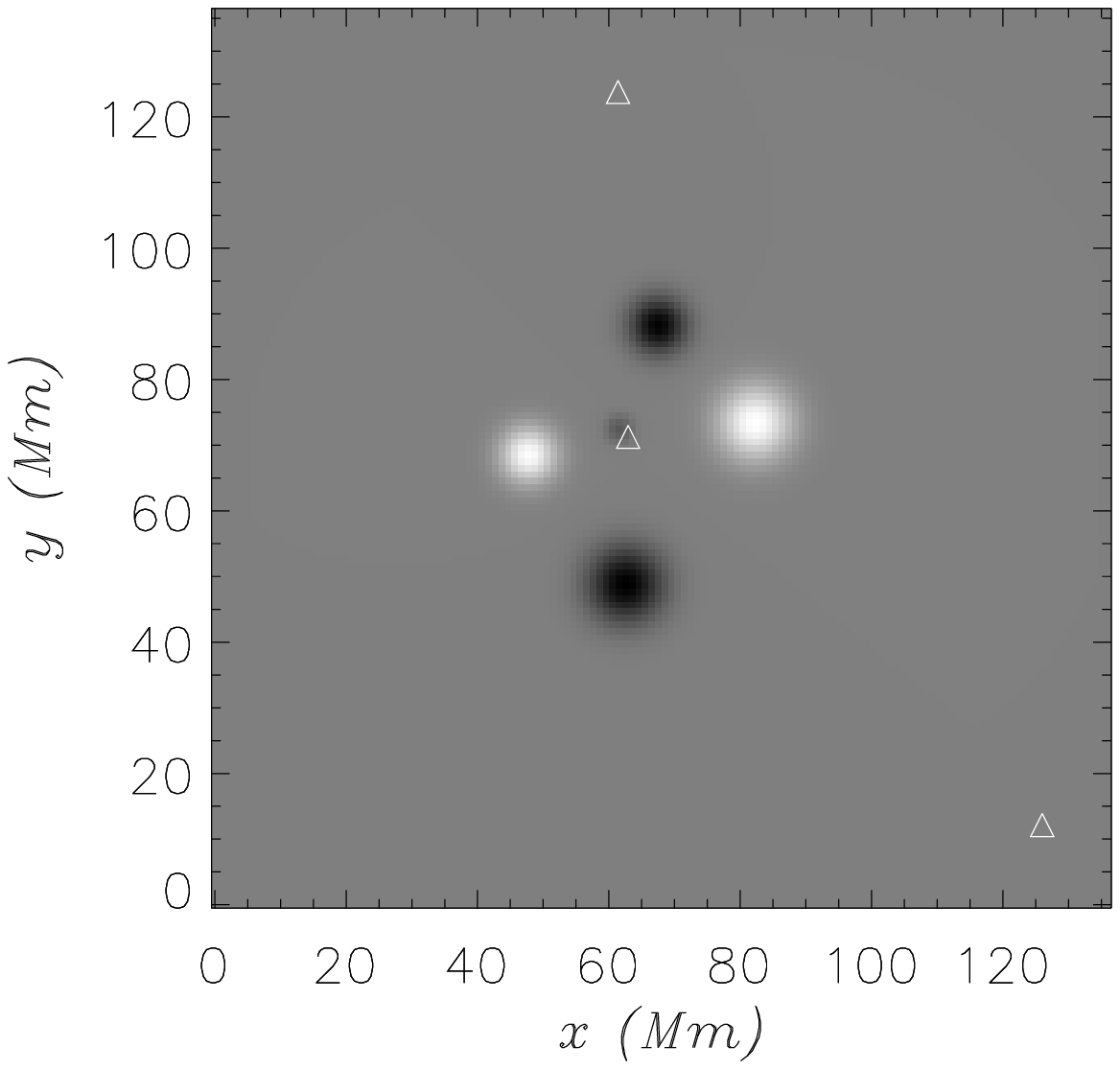}
\includegraphics[width=.49\textwidth]{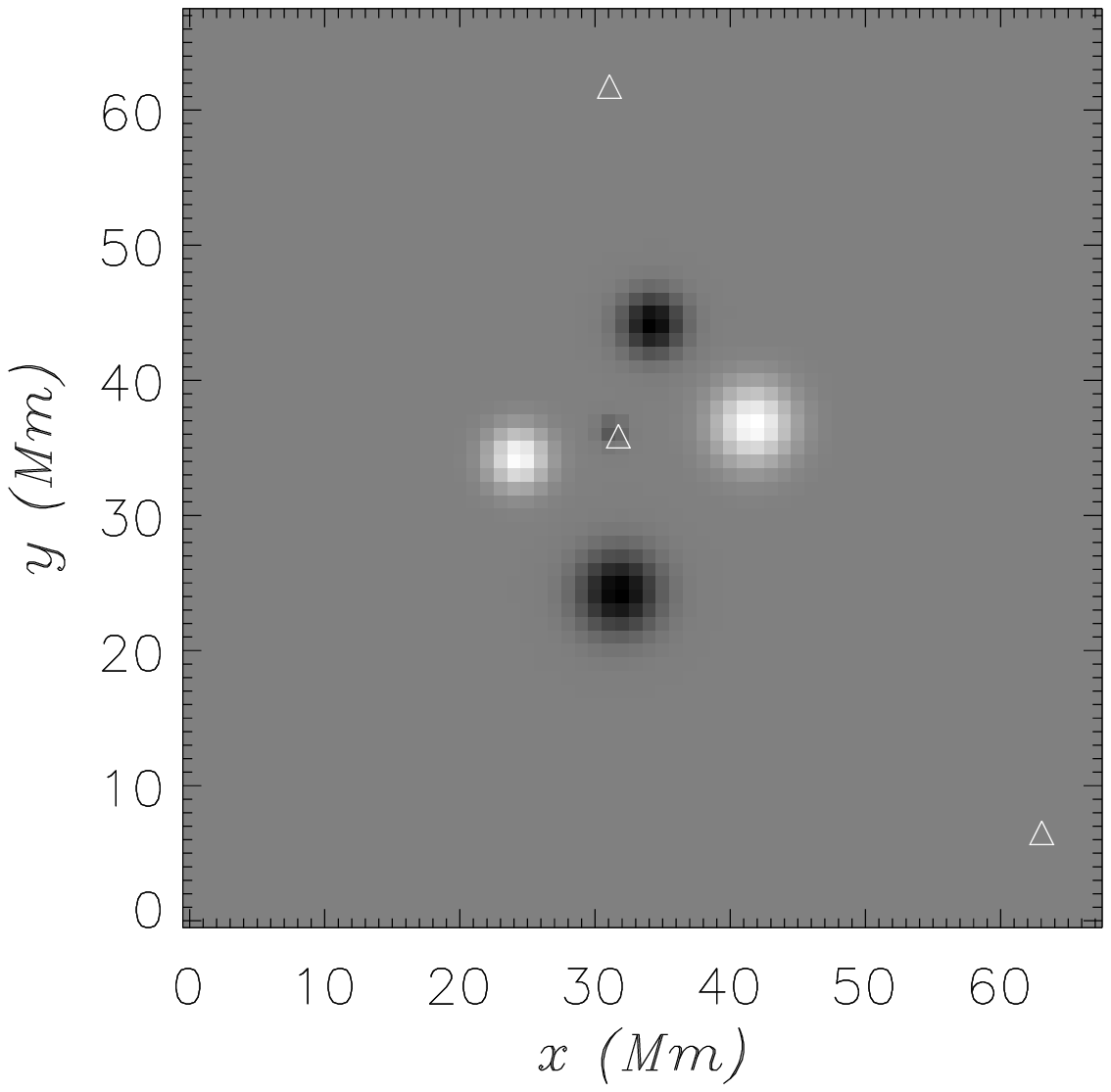}
\includegraphics[width=.49\textwidth]{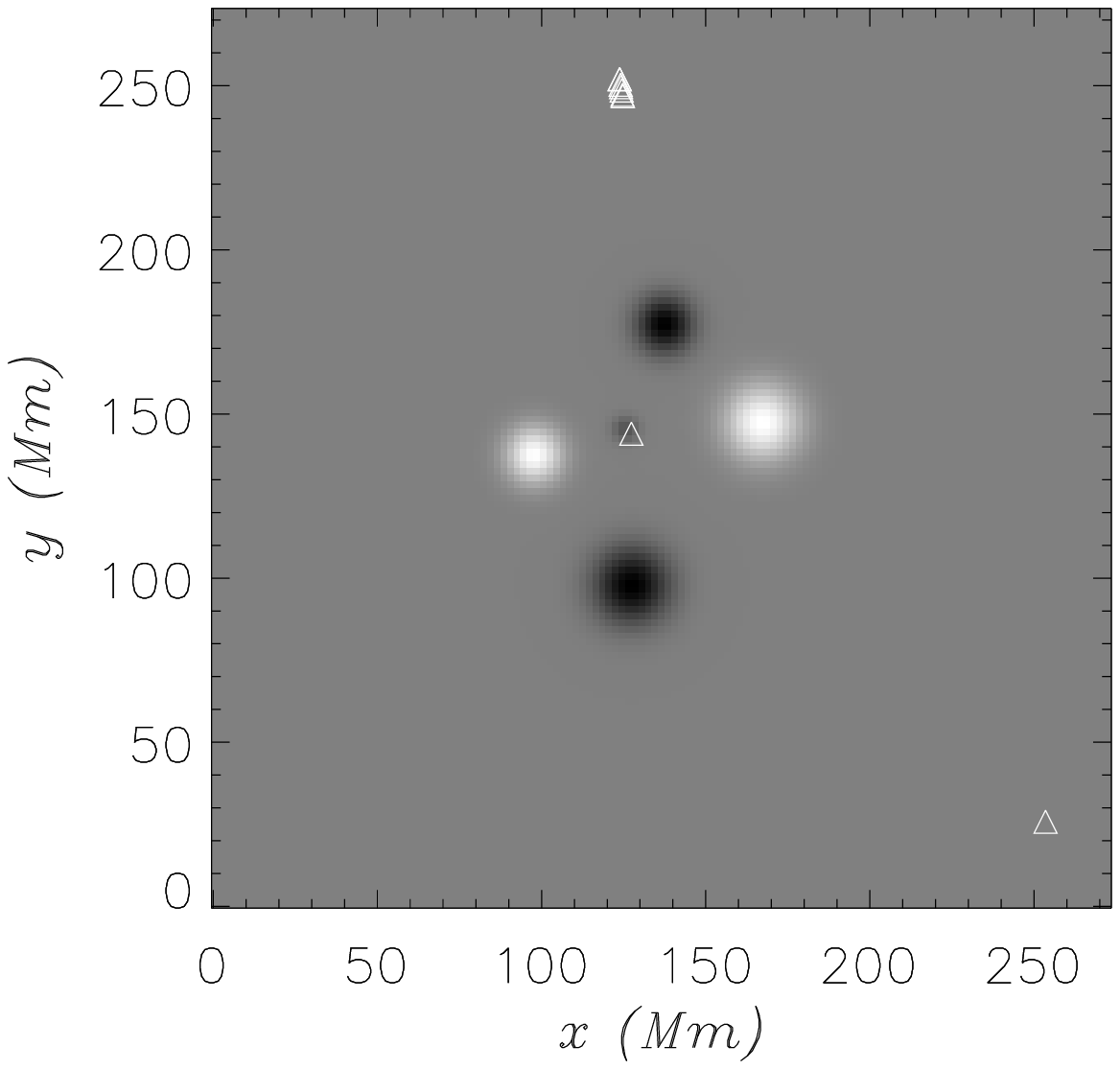}
\includegraphics[width=.49\textwidth]{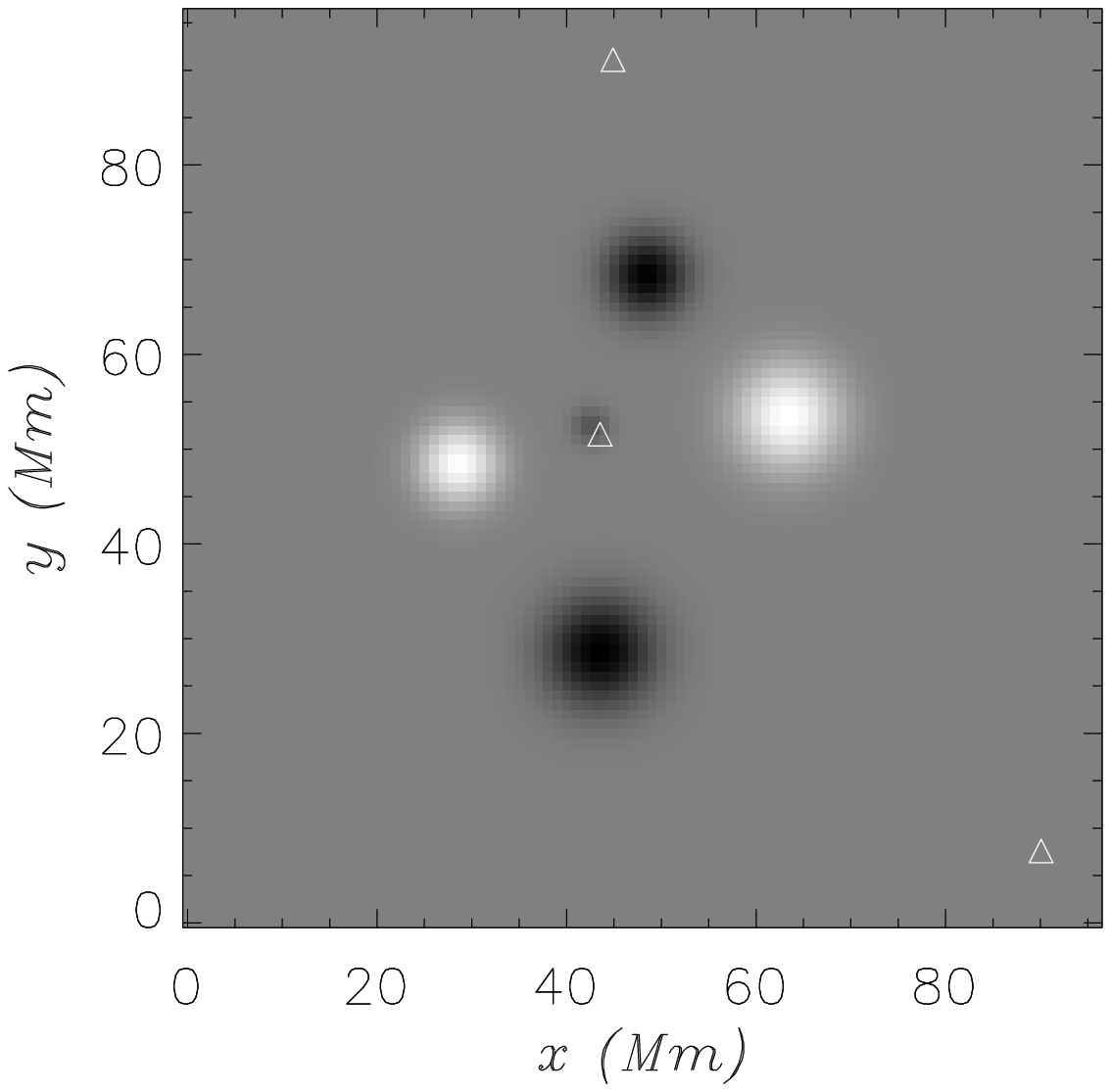}
\includegraphics[width=.49\textwidth]{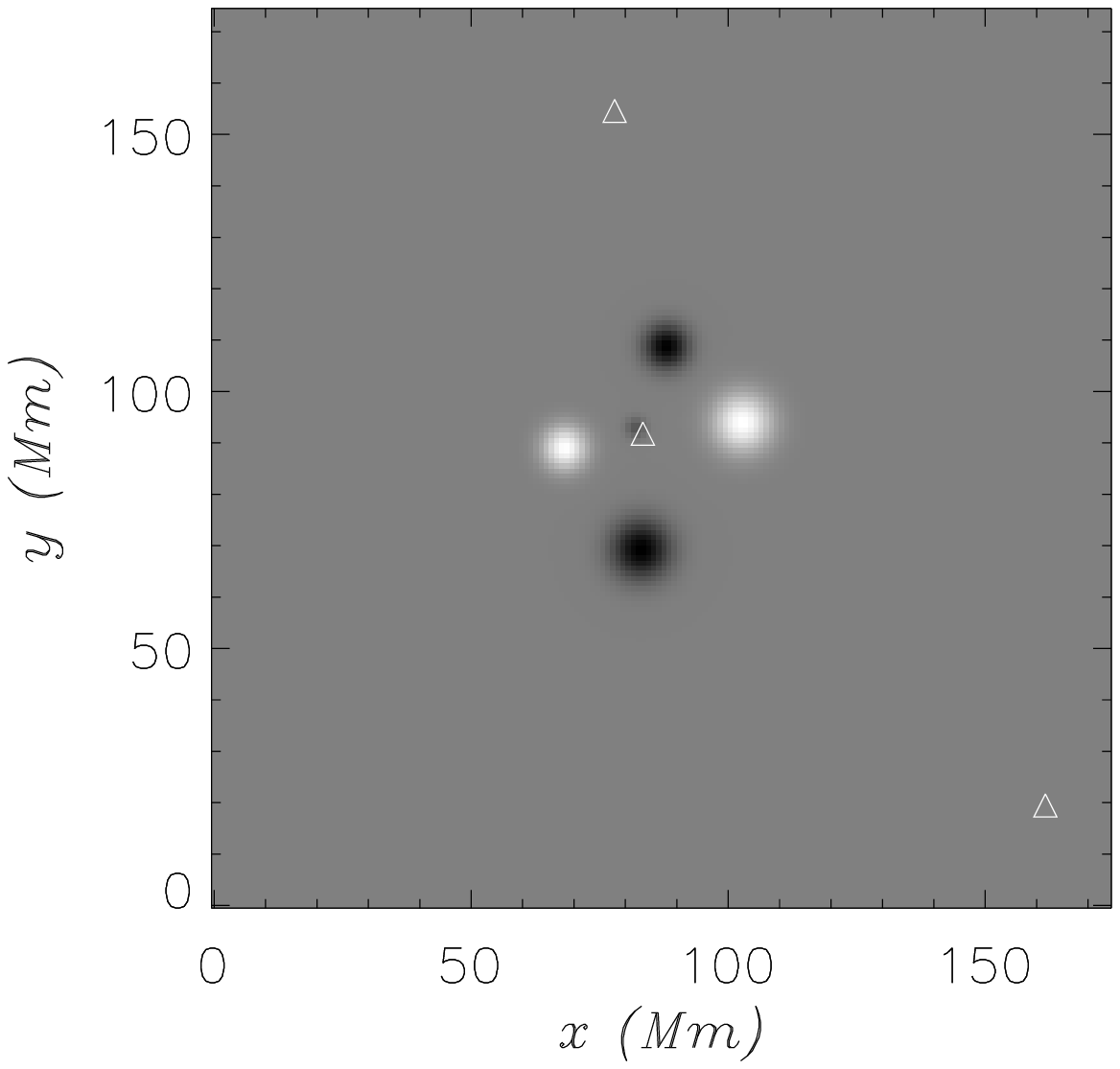}
\caption{Location of null points (triangles) on the {\em xy}-plane for different
potential fields. Top row: potential fields with different boundary conditions:
(Left) open sides and top boundaries, (Right) periodic. Middle row: potential
fields with a different spatial resolution for the same field of view: (Left)
decreased by a factor of two, (Right) increased by a factor of two. Bottom row: potential
fields for different field-of-view: (Left) reduced by 20 pixels on each edge,
(Right) increased by 20 pixels on each edge.}
\label{fig:np_pot}
\end{figure}

The above computation were performed using a Grad--Rubin numerical scheme and
assuming closed boundary conditions on the sides and top of the computational
box. We now compute the magnetic configurations for a potential field with open
boundary conditions and with periodic boundary conditions. In
Figure~\ref{fig:np_pot} top row, we plot the location of the null points for these
new boundary conditions. This has to be compared to Figure~\ref{fig:np_pot_loc}.
The location of NP0 is similar for all models and the spectral radius of NP0 is
again the strongest (see Table~\ref{tab:prop1}). The potential field with
periodic boundary conditions has created two negative null points near the sides
of the computational box. The topology of the potential field is slightly
influenced by the boundary conditions for this experiment. 

\begin{table}
\caption{Properties of null points (type, location, eigenvalues) for the
different potential fields}
\label{tab:prop1}
\begin{tabular}{lcccc}
\hline \\[-0.4cm]
Model & Type & Location & Eigenvalues \\
 & & ($x_0$, $y_0$, $z_0$) & ($\lambda_1 = \pm \rho_J$, $\lambda_2$,
 	$\lambda_3$)  \\
\hline \\[-0.2cm]

Boundary Conditions & & & \\[0.2cm]
{\em Closed Boundaries} & + & (64.30, 72.94, 6.90) & 
	({\bf--0.026}, 0.021, 5.5 10$^{-3}$) \\[0.2cm]

{\em Open Boundaries} & + & (63.3, 71.96, 7.39) & 
	({\bf--0.019}, 0.016, 5.2 10$^{-3}$) \\[0.2cm]

{\em Periodic Boundaries} & + & (62.93, 71.25, 7.91) & 
	({\bf--0.023}, 0.017, 5.5 10$^{-3}$) \\
& -- & (61.4, 123.8, 0.02) & 
	(8.6, --5.1, --3.5)$\times$ 10$^{-3}$ \\
& -- & (126, 12.15, 5.38) & 
	(2.3, 0.02, 2.28)$\times$ 10$^{-4}$ \\[0.2cm]
	
\hline \\[-0.2cm]
Spatial Resolution & & & \\[0.2cm]
{\em 70$\times$70$\times$60}  & + &  (31.7, 35.8, 3.91) & 
	 ({\bf--0.047}, 0.035, 0.01) \\
& -- & (31.06, 61.75, 0.01) & 
	(1.9, --1.1, --0.7)$\times$ 10$^{-3}$ \\[0.2cm]
& -- & (63.05, 6.51, 3.14) & 
	(4.73, --4.68, --0.04)$\times$ 10$^{-4}$ \\[0.2cm]

{\em 280$\times$280$\times$240} & + & (127.37, 144, 15.84) & 
	 ({\bf--0.011}, 2.8 10$^{-3}$, 8.7 10$^{-3}$) \\
 & -- & (124.74, 246.8, 0.022) & 
	(4.7, --2.76, --1.92)$\times$ 10$^{-4}$ \\
 & -- & (124.66, 247.22, 0.022) & 
 	(4.6, --2.72, --1.9)$\times$ 10$^{-4}$ \\
 & -- & (124.4, 248.57, 0.023) & 
	(4.4, --2.58, --1.81)$\times$ 10$^{-4}$ \\
 & -- & (124.22, 249.57, 0.024) & 
	(4.25, --2.5, --1.75)$\times$ 10$^{-4}$ \\
 & -- & (124.08, 250.32, 0.024) & 
	(4.15, --2.44, --1.71)$\times$ 10$^{-4}$ \\
 & -- & (123.8, 251.91, 0.025) & 
	(3.97, --2.32, --1.64)$\times$ 10$^{-4}$ \\
 & -- & (123.77, 252.06, 0.025) & 
	(3.96, --2.31, --1.63)$\times$ 10$^{-4}$ \\
 & -- & (253.49, 25.76, 8.39) & 
	(1.15, --1.16, --0.007)$\times$ 10$^{-4}$ \\[0.2cm]

\hline \\[-0.2cm]
Field-of-View & & & \\[0.2cm]
{\em 100$\times$100} & + &  (43.5, 51.59, 7.83) & 
	({\bf--0.024}, 0.017, 5.7 10$^{-3}$) \\
 & -- &  (44.9, 91.1, 0.18) & 
	(5.5, --2.9, --2.6)$\times$ 10$^{-3}$ \\
 & -- &  (90.07, 7.59, 14.8) & 
	(8.8, --8.4, --0.48)$\times$ 10$^{-4}$ \\[0.2cm]
	
{\em 180$\times$180} & + &  (83.41, 91.8, 7.92) & 
 	({\bf --0.023}, 0.017, 5.5 10$^{-3}$) \\
 & -- & (77.9, 154.57, 0.025) & 
	(2.6, --1.64, --0.99)$\times$ 10$^{-4}$ \\
 & -- & (161.7, 19.45, 1.16) & 
	(7.93, --7.72, --2.18)$\times$ 10$^{-5}$ \\[0.2cm]
\hline
\end{tabular}
\end{table}

\subsection{Effect of Spatial Resolution}

To study the effects of the spatial resolution on the topology of potential
fields, we modify the pixel size by a factor of 0.5 and 2
(Figure~\ref{fig:np_pot} middle row). We also make sure that the total unsigned
magnetic flux remains unchanged (variation less than 1\%). The computations are
carried out with periodic boundary conditions. NP0 is located nearly at the same
location for both spatial resolutions. The other null points are similar to
those appearing in the potential field with periodic conditions. The cluster of
eight null points suggests that there is a null line at this location. The topology
of the potential field is thus similar to the topology of the potential field
with closed boundary conditions. We again emphasize that the spectral radius of
NP0 remains almost constant and is the strongest (see Table~\ref{tab:prop1}).

\subsection{Effect of Field-of-View}

We now modify the field-of-view of the experiment but keeping the same spatial
resolution and the same total unsigned magnetic flux. In Figure~\ref{fig:np_pot}
bottom row, we plot the location of the null points for two different
fields-of-view using a potential field with periodic conditions: we decrease or
increase the number of pixels by 40 pixels in each direction (20 pixels on each
edge). The properties of the null points are summarised in
Table~\ref{tab:prop1}. The number of null points is the same for the different
fields-of-view. The spectral radius of NP0 is similar and still the strongest.
The topology of the magnetic configuration is almost not influenced by the
field-of-view for this experiment. It is worth noticing that the field-of-view
was modified in such way that there is no magnetic flux on the edges of the
computational box as this would strongly influence the magnetic configuration
and also that it would have unexpected effects due to the violation of the
solenoidal condition.

\end{article} 
\end{document}